\documentclass[12pt]{article}

\pdfoutput=1

\usepackage{graphicx}
\usepackage{caption}
\usepackage{subcaption}
\usepackage{authblk}

\usepackage[hyphens,spaces,obeyspaces]{url}
\usepackage[hidelinks]{hyperref}

\newcommand{\um}{$\mu m$}
\newcommand{\pitch}[2]{${#1}\ \mu m \times {#2}\ \mu m$}
\newcommand{\Neq}{$n_{eq}/cm^{2}$}
\newcommand{\degr}{$^\circ$}

\begin{document}

\title{First study of small-cell 3D Silicon Pixel Detectors for the High Luminosity LHC}

\author[1]{E. Curr\'{a}s}
\author[1]{J. Duarte-Campderr\'{o}s}
\author[1]{M. Fern\'{a}ndez}
\author[1]{A. Garc\'{\i}a}
\author[1]{G. G\'{o}mez}
\author[1]{J. Gonz\'{a}lez}
\author[1]{R. Jaramillo}
\author[1]{D. Moya}
\author[1]{I. Vila}
\author[2]{S. Hidalgo}
\author[2]{M. Manna}
\author[2]{G. Pellegrini}
\author[2]{D. Quirion}
\author[3]{D. Pitzl}
\author[4]{A. Ebrahimi}
\author[5]{T. Rohe}
\author[5]{S. Wiederkehr}

\affil[1]{Instituto de F\'{\i}sca  de Cantabria, University of Cantabria / CSIC.}
\affil[2]{Instituto de Microelectr\'{o}nica de Barcelona, Centro Nacional de Microelectr\'{o}nica.}
\affil[3]{Deutsches Elektronen Synchotron (DESY).}
\affil[4]{University of Hamburg.}
\affil[5]{Paul Scherrer Institut.}

\date{} 
\setcounter{Maxaffil}{0}
\renewcommand\Affilfont{\itshape\small}

\maketitle

\begin{abstract}
A study of 3D pixel sensors of cell size \pitch{50}{50} fabricated at IMB-CNM using double-sided n-on-p 3D technology is presented. 
Sensors were bump-bonded to the ROC4SENS readout chip. For the first time in such a small-pitch hybrid assembly, the 
sensor response to ionizing radiation in a test beam of 5.6 GeV electrons was studied. 
Results for non-irradiated sensors 
are presented, including efficiency, charge sharing, signal-to-noise, and resolution for different incidence angles. 

\end{abstract}

\section{Introduction}
The LHC accelerator at CERN~\cite{ref:LHC} will undergo a high luminosity upgrade, currently planned for the years 2023-2024,
after which the instantaneous luminosity will reach $10^{35} cm^{-2} s^{-1}$ and the average number of hard proton-proton
scatters during any single proton bunch crossing (pileup) will exceed 140. This will require upgraded detectors
able to cope with extremely high track densities and to withstand unprecedented levels of radiation. In particular, vertex
detectors must be tolerant to hadron fluences up to $2 \times 10^{16}$ \Neq, so detectors with a small charge carrier drift
distance and a high electric field close to the collecting electrodes are required to minimize the impact of radiation-induced 
trapping. For a typical planar pixel sensor, this implies
reducing as much as possible the sensor thickness. One possible choice of technology for the innermost layers, where the
levels of radiation are expected to be highest, is 3D pixels, in which n-type columns are etched perpendicularly into a p-type bulk.
Figure~\ref{fig:3Dsketch} shows a cross-section of such a device. With this arrangement of electrodes, charge carriers travel
parallel to the sensor surface, and their travel distance is decoupled from the sensor thickness. For small pixel sizes,
the sensors are intrinsically resistant to radiation and can be fully depleted at very low bias voltages owing to the
small inter-electrode distance. Pixel sensors of very small pixel size are required to achieve a reasonably low detector
occupancy. We present the first study of 3D pixels of cell size \pitch{50}{50} bump-bonded to a readout chip of matching
pixel cell size. 


\section{Experimental setup}

\subsection{Sensors and readout}

\begin{figure}
    \centering
    \begin{subfigure}{.6\textwidth} 
      \includegraphics[width=\textwidth]{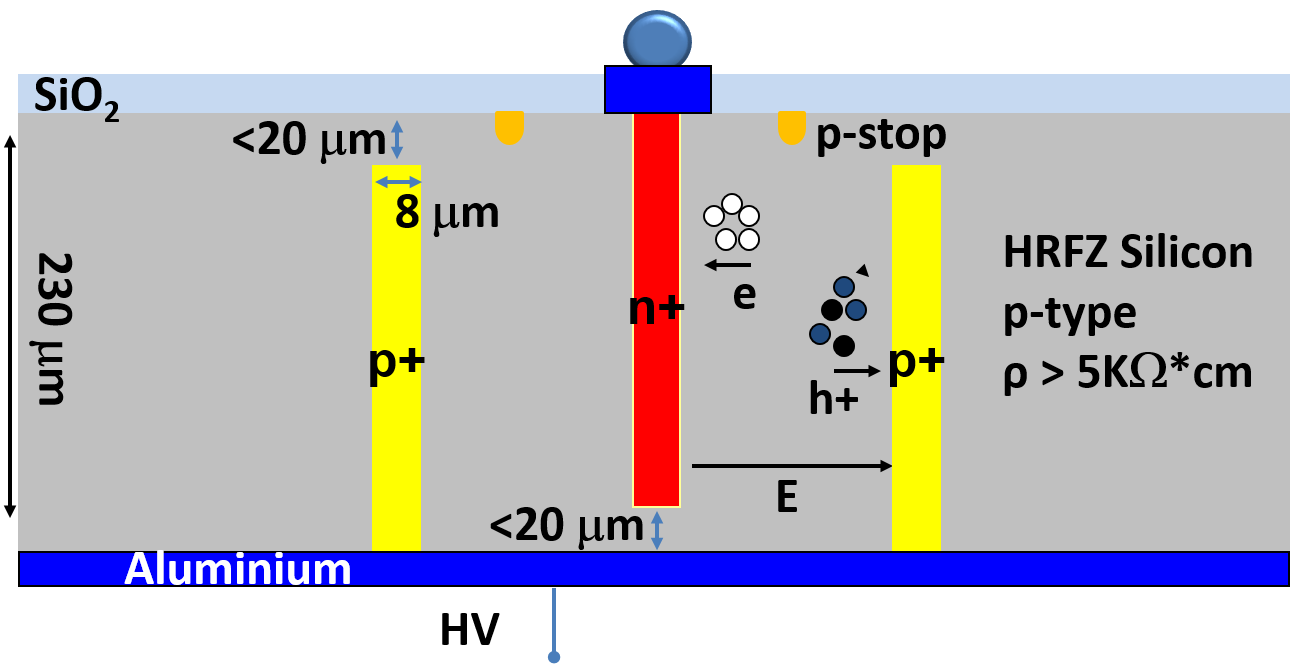}
  \end{subfigure}
    \hfill
  \begin{subfigure}{.3\textwidth}
      \includegraphics[width=\textwidth]{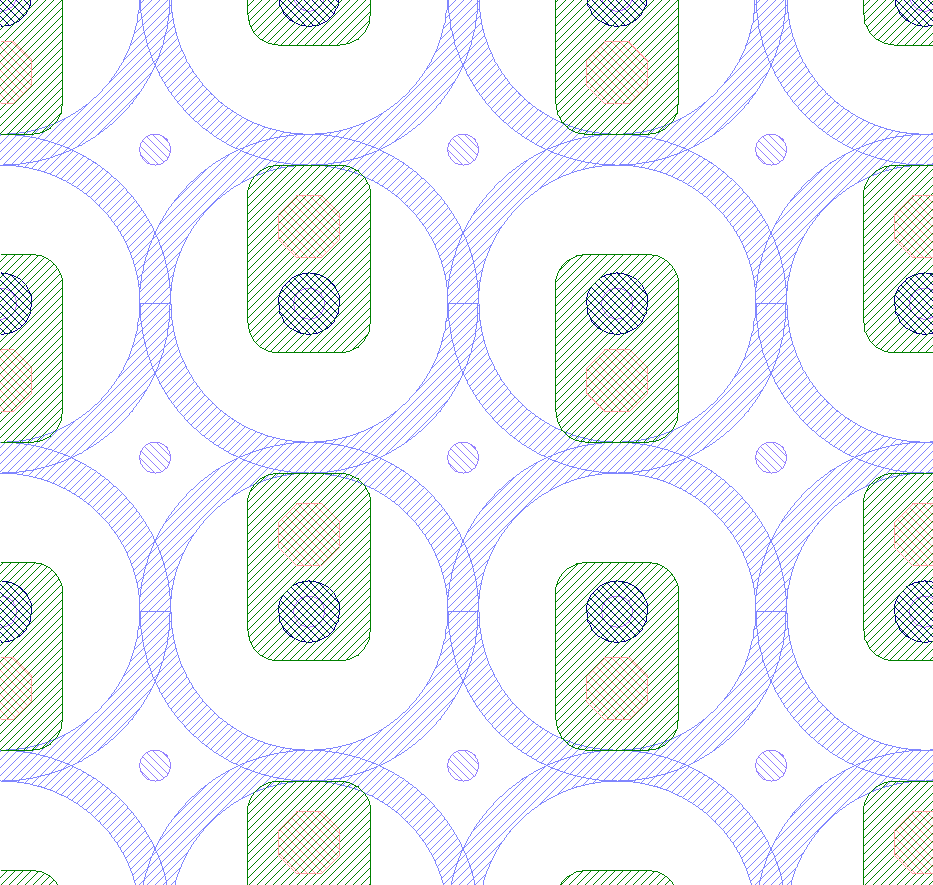}
  \end{subfigure} 
  \caption{Left: cross-section of a double-sided 3D pixel sensor. Right: a zoomed view of the pixel sensor layout.
           The n-columns at the center of the p-stop rings are etched from the top side. The p-columns are etched from the bottom side.}
  \label{fig:3Dsketch}
\end{figure}

The sensors studied were fabricated at IMB-CNM using an n-on-p double-sided technology~\cite{ref:3Dpix}, where n-type (junction) columns
are etched from the ``top'' side into a p-bulk, and p-type (ohmic) columns are etched from the ``bottom'' side. Columns have a diameter of 
8 \um\ and are not fully passing. The entire sensor has a thickness of 230 \um, and the pixel cell size is \pitch{50}{50}. 
A p-stop ring surrounds each junction column to isolate the electrodes from accumulated surface charges. 
Figure~\ref{fig:3Dsketch} shows a sketch of the transverse cross-section (left) and a zoomed view of the layout (right) of such a sensor.
Each sensor has a matrix of 155 columns $\times$ 160 rows for a total of 24800 pixels. The sensors have a Ni/Au underbump metallization (UBM) 
deposited 
through an electroless process at CNM, and were bump-bonded to a front-end readout electronics called ROC4SENS~\cite{ref:R4S} fabricated using a 
250 nm CMOS process with radiation tolerant design. The ROC4SENS is a generic pixel readout 
chip (ROC) developed by the Paul Sherrer Institut (PSI)~\cite{ref:PSI} specifically to characterize the sensor part of a hybrid pixel detector. 
In order not be limited to signals exceeding a certain threshold the chip has no discriminator for zero suppression. This in return
means that the sampling time has to be defined by an external trigger (hold signal). After triggering the chip is insensitive to further
signals until the hold signal is released after readout is finished. This dead  time depends on the readout speed of the DAQ system and is
not problematic for the application in test beams. In order to keep design and operation simple, no DACs have been implemented in the chip 
and four reference voltage have to be provided externally.

The single chip assembly (SCA) consisting of a sensor bump bonded to a R4S chip is glued and wire bonded to a dedicated printed circuit
board (PCB) which is connected via a special interface to a so called digital test board (DTB) originally developed by PSI for the testing
of the digital CMS pixel chip. In this paper we report on tests with 2 SCAs called SCA ``A'' and ``B''.

\begin{figure}
    \centering
    \begin{subfigure}[b]{.49\textwidth}
        \includegraphics[width=\textwidth]{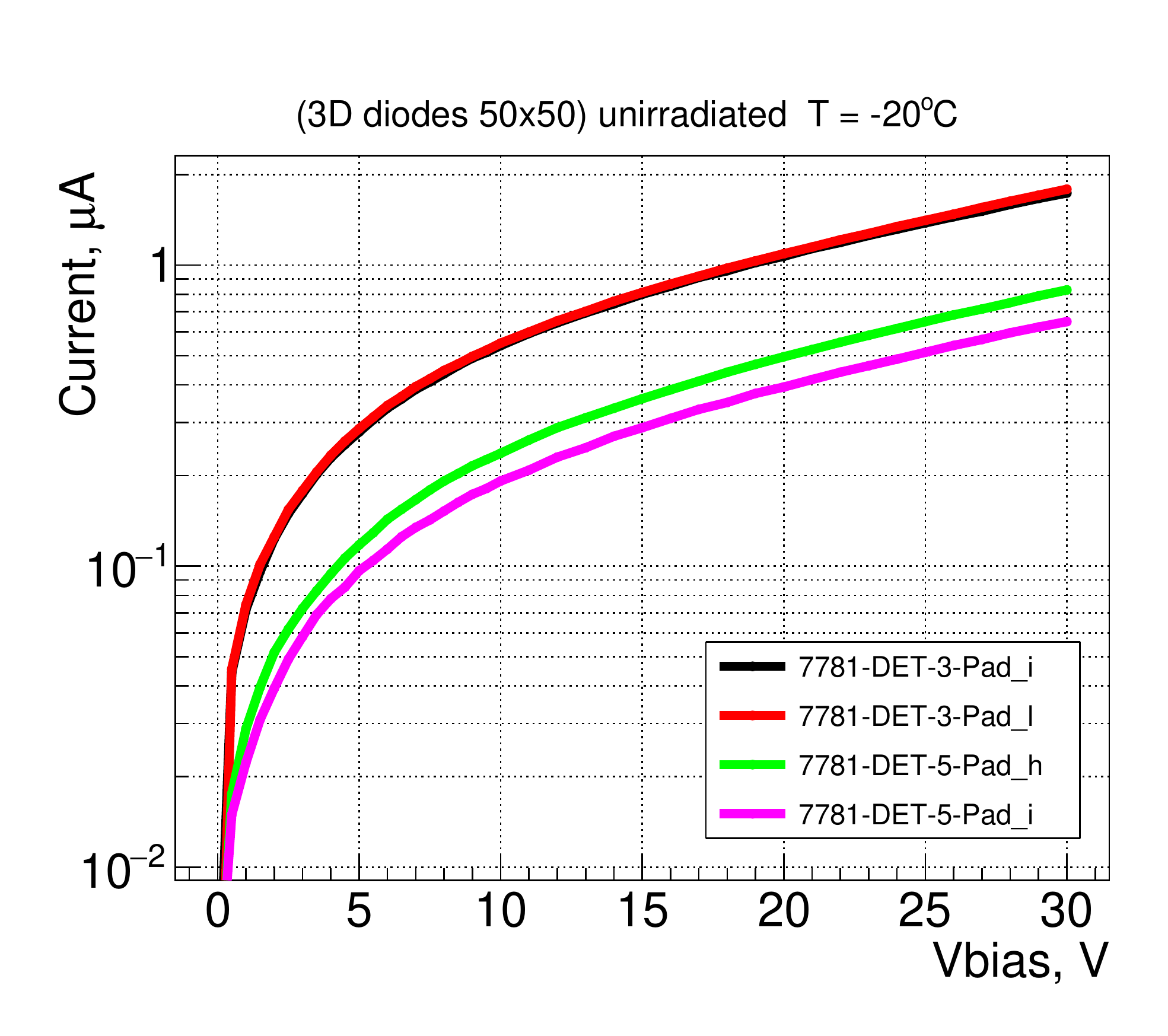}
    \end{subfigure} 
    \begin{subfigure}[b]{.49\textwidth}
        \includegraphics[width=\linewidth]{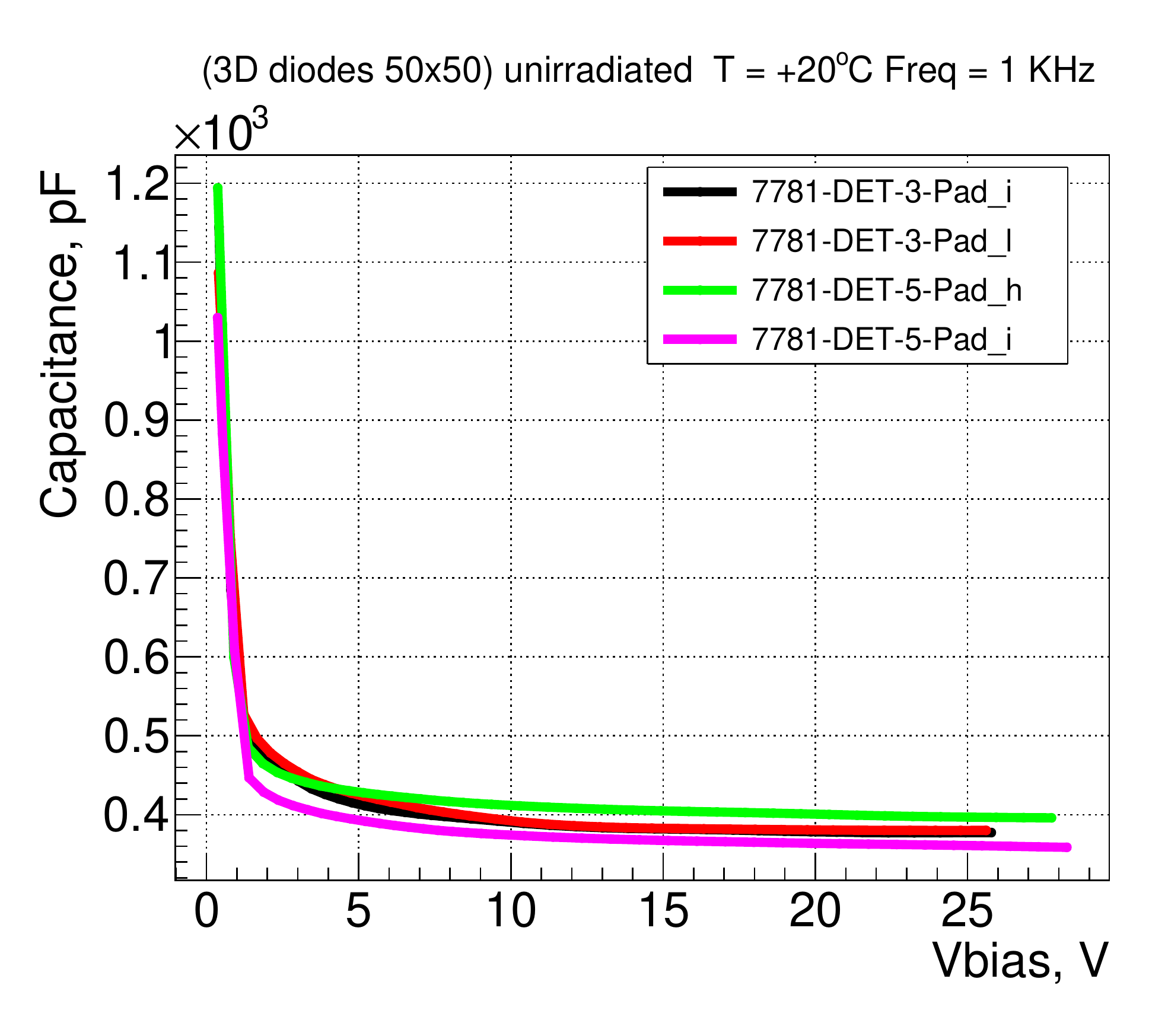}
    \end{subfigure}
    \caption{Electrical properties of small-pitch 3D pixels in pad-like sensors. 
    Left: Total dark current as a function of reverse bias voltage. Right: Total capacitance as a function of reverese bias voltage.}
    \label{fig:PADIVCV}
\end{figure}

In order to study the electrical characteristics of 3D pixels of this cell size, small pad-like sensors were produced in the same wafers. These sensors
contain only 100 $\times$ 100 3D pixels in which all electrodes (n-type columns) are connected together via an Aluminum metallization, but which
are otherwise identical to the pixel cells previously described. 
Figure~\ref{fig:PADIVCV} shows the total dark current (left) and capacitance (right) as a function of reverse bias voltage for four of these pad-like 
sensors.
Taking into account the fact that each pad sensor contains 10000 pixels, these figures show that, at a voltage of 20 V and a temperature of 
-20\degr\ C, the dark current per pixel varies from 40 to 100 pA. The per-pixel capacitance is found to be about 40 fF, and the depletion voltage is 
about 10 V, considerably lower than for typical planar pixel sensors. 

\subsection{Test Beam Setup}

\begin{figure}
  \centering
    \includegraphics[width=0.9\textwidth]{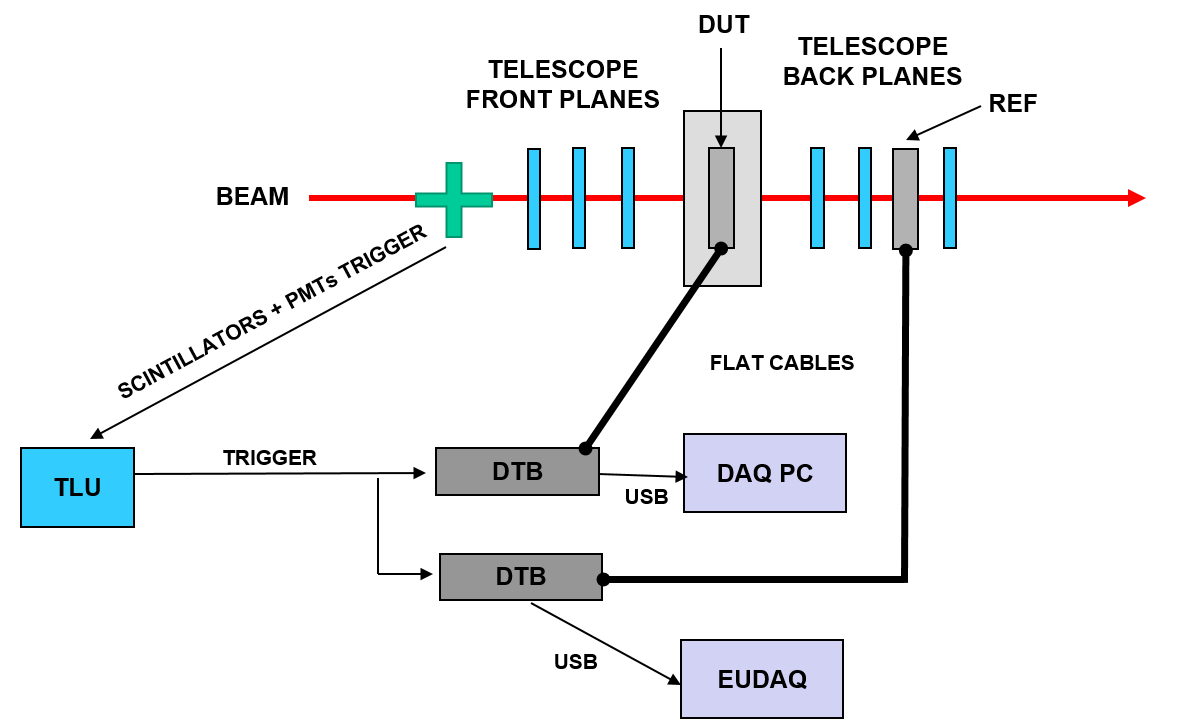}
  \caption{Sketch of test beam setup.}
  \label{fig:tb-setup}
\end{figure}

The sensor response to energetic minimum ionizing particles was studied at the DESY test beam facility~\cite{ref:DESY}, which provides a beam of 
electrons of energy in the range 1 -- 6 GeV adjustable through a spectrometer dipole magnet. The DESY synchrotron circulates a single electron
bunch at a frequency of 1 MHz, and the particle rate in the test beam depends on beam line, energy, collimator settings, target material and 
operation. For the study presented here, an energy of 5.6 GeV was used and particle rates of about 40 kHz were achieved.  
Tracking was provided by the DATURA beam telescope~\cite{ref:telescope}, which consists of six planes of MIMOSA26 pixel detectors of pitch 
\pitch{18.4}{18.4}, with an intrinsic telescope plane resolution of about 3.2 \um\ in the best possible conditions of particle energy, scattering and
distance between planes~\cite{ref:telres}. A trigger logic unit (TLU)~\cite{ref:TLU} requires coincidence of two 
trigger input signals generated by a pairs of PMT-scintillator assemblies placed upstream of the telescope, forming a trigger window of approximately 
1 cm $\times$ 1 cm, slightly larger than the device under test (DUT). The DUT is placed in the middle of the telescope, sandwiched by two triplets 
of telescope planes. The sensor mother board is mounted on a motorized platform which allows to remotely adjust the position and the inclination 
of the sensor with respect to the beam.
The MIMOSA sensors operate in a rolling shutter mode with an integration time of approximately 120 $\mu$s, which causes track pile-up. 
A SCA with pitch \pitch{100}{150} equipped with a fast PSI46V2.1dig chip, as used in the outer layersof the CMS Phase-1 upgrade pixel detector, 
is used as timing reference to tag the track on the DUT. It is also read out with a DTB.
%
%
Figure~\ref{fig:tb-setup} shows a schematic diagram of the test beam setup.

\section{Data reconstruction}

Data from the telescope planes and the reference sensor are stored via specialized producers by the EUDAQ data acquisition framework~\cite{ref:EUDAQ}.
Data from the DUT (ROC4SENS), not yet integrated into EUDAQ, was pre-processed by the digital test board (DTB) with a customized firmware and stored via USB
on a separate data acquisition PC. These two data streams share the same TLU trigger, and are fed into an analysis software which integrates all 
information and performs an offline event reconstruction. 

\begin{figure}
  \centering
    \includegraphics[height=0.5\textwidth,width=0.55\textwidth]{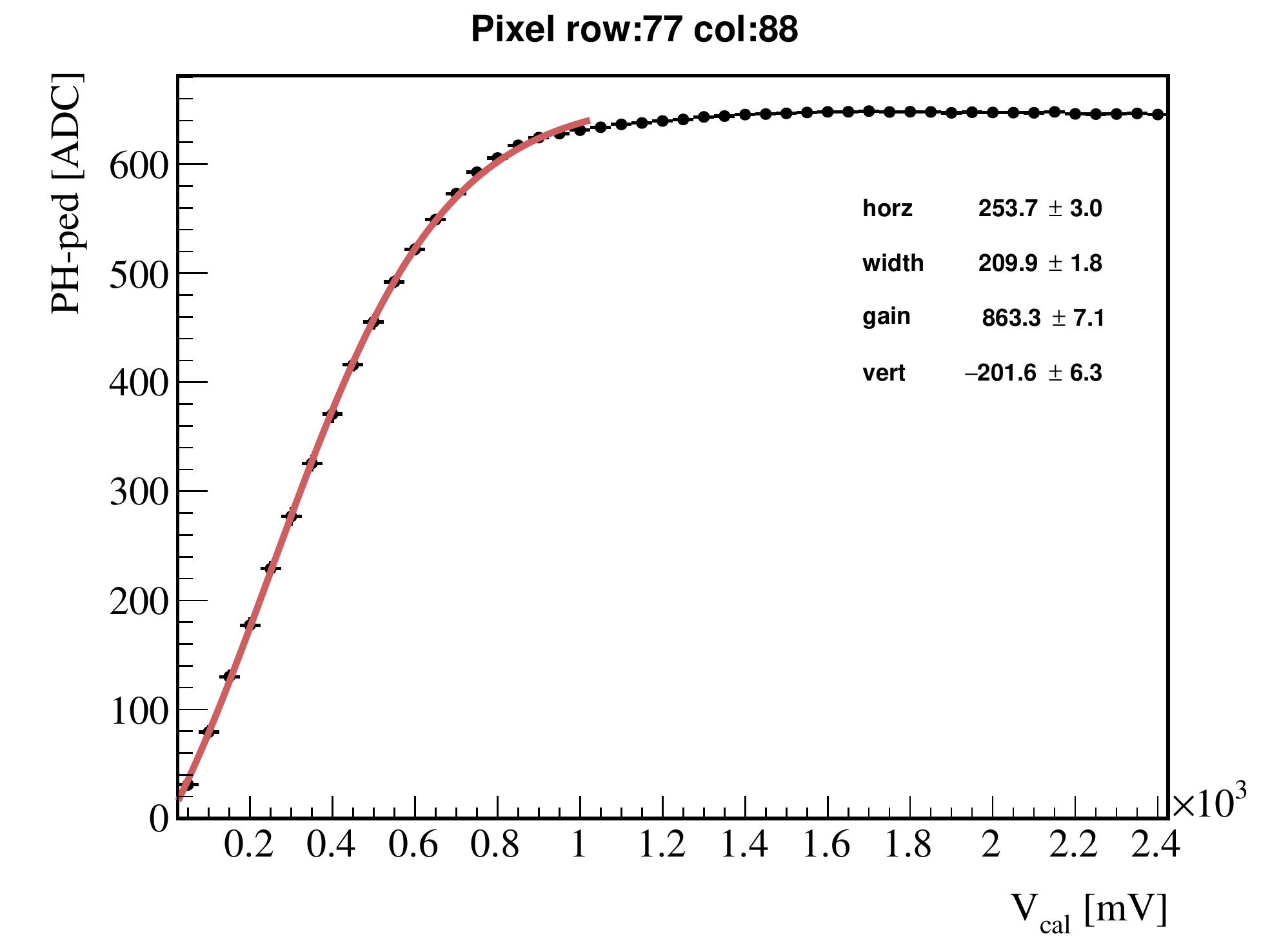}
  \caption{Example of calibration for one single ROC pixel. The pulse height is recorded (in units of ADC) as a function of an applied calibration voltage
          (Vcal, in mV). The fit is then used to obtain the gain of each ROC pixel.}
  \label{fig:calib}
\end{figure}

A calibration of each pixel of the readout chip is performed by injecting a calibration pulse.
Figure~\ref{fig:calib} shows the calibration for one single ROC pixel, where the pedestal-subtracted pulse height in units of ADC is recorded as a function of the 
calibration signal amplitude (Vcal). A Fermi function is fit to these distributions and applied offline to each pixel separately to properly handle 
relative pixel-to-pixel gain variations.

\begin{figure}
  \centering
    \includegraphics[width=0.6\textwidth]{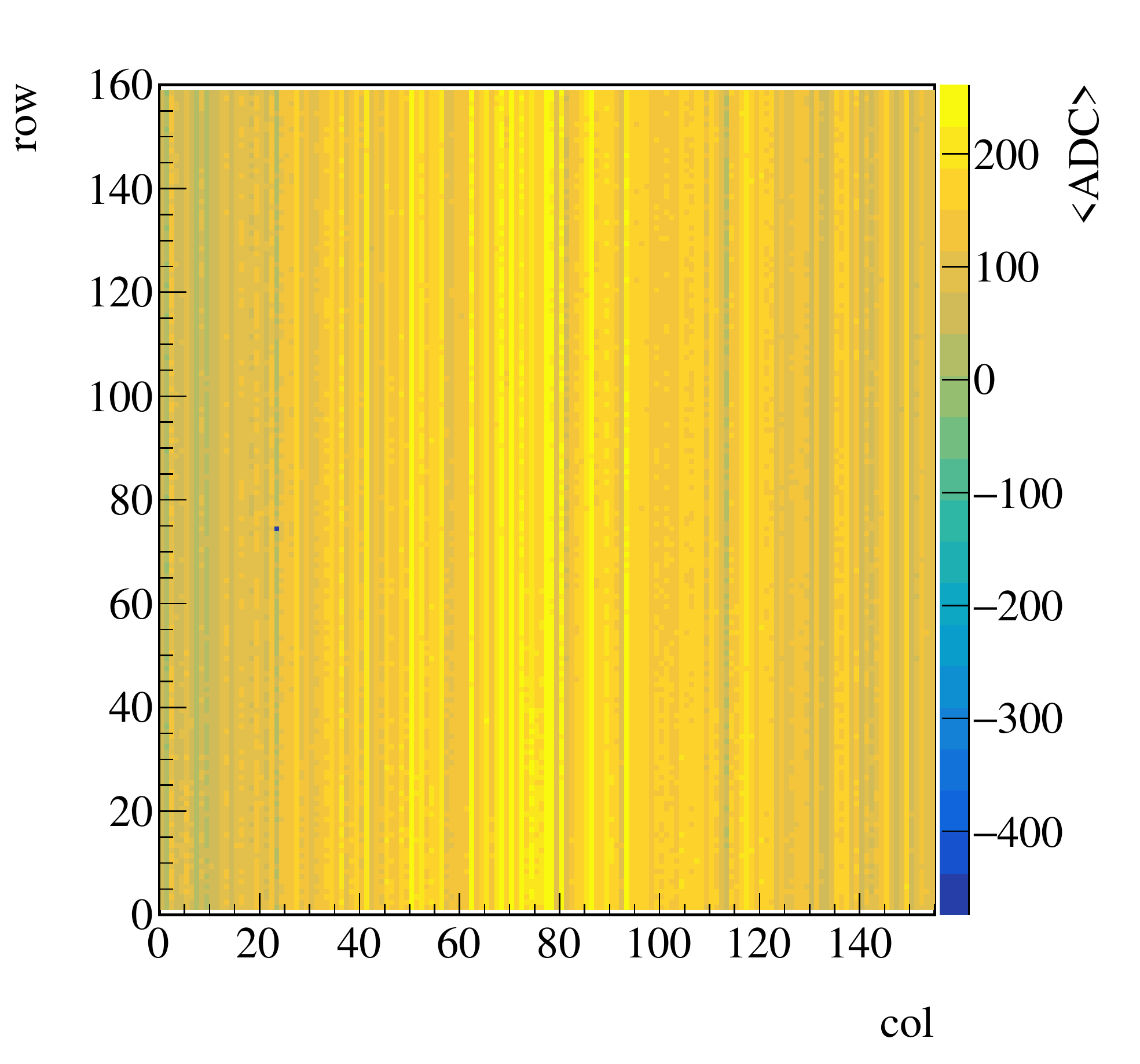}
  \caption{Pedestal map. Average pedestal for each pixel in ADC counts.}
  \label{fig:pedestalmap}
\end{figure}

\begin{figure}[!hbt]
  \centering
    \includegraphics[height=0.55\textwidth,width=0.55\textwidth]{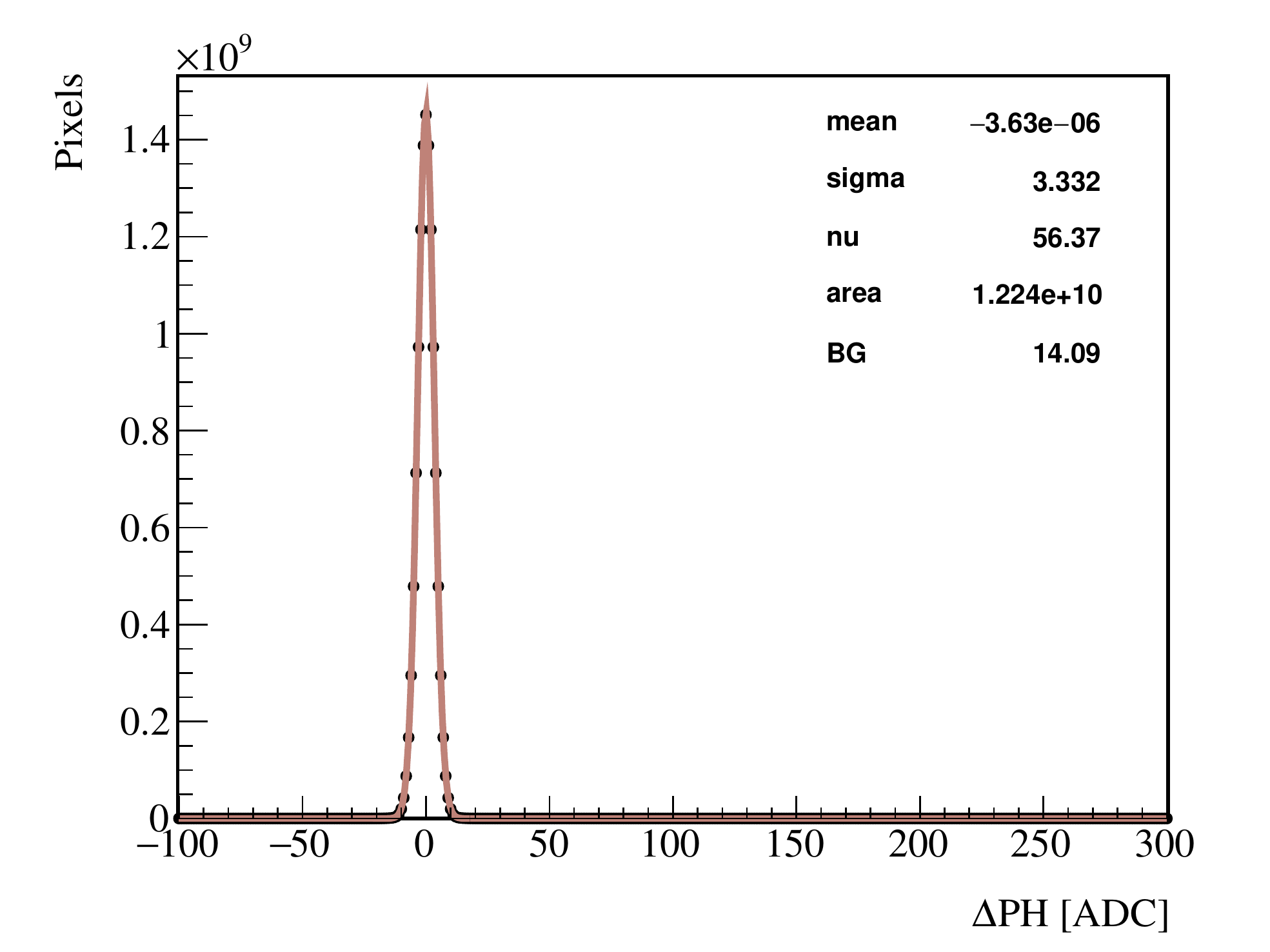}
  \caption{Pedestal-subtracted pulse height for all pixels in units of ADC counts. A t-Student function is fit to the distribution. 
           The sigma parameter of the fit provides an estimate of the ROC noise.}
  \label{fig:noisegauss}
\end{figure}

Once the data taking begins, the DTB uses the first 100 readouts to determine the pedestal of each pixel.
Figure~\ref{fig:pedestalmap} shows the average pedestal value (in ADC counts) recorded for each pixel. The column pattern is a known ROC feature which arises 
from gyrator circuit variations at the end of each column.   
During a run the pedestal subtracted pulse height ($\Delta PH$) is calculated for each pixel. The difference between neighbouring pixels is used to define if a pixel was hit, 
with a threshold value of 24 ADC counts. A region of interest (ROI) of 7 rows times 5 columns around the hit pixel is stored for off line processing. This ROI technique is 
used to reduce the amount of data, as there are limits in the memory and data transfer of the DTB.  
Hits are therefore defined based on a relative signal threshold and not on a signal-to-noise cut. 
The noise, calculated as the width of the $\Delta PH$ distribution, is approximately 3 ADC counts, as shown in Figure~\ref{fig:noisegauss}. 
Offline, a column-wise common mode correction is applied for each pixel in the ROI by subtracting the nearest edge pixel in each of the 5 columns to in-between 
pixels. For the corrected pixels, pedestal-subtracted pulse heights are recalculated, and hits are re-defined by requiring a difference greater than 10 ADC counts
with respect to neighboring pixels. Finally, all neighboring hits in rows and columns are clustered, and the calibration Fermi functions are applied to each pixel to
obtain a uniform charge response.

The calibrated charge obtained is normalilzed to the expected average charge deposition of 76 electrons per micron for a MIP 
traversing an unirradiated sensor, since the equivalence between charge and voltage for the calibration pulse of the ROC4SENS is not well known. 

\section{Results}
\begin{figure}[!hbt]
  \centering
    \includegraphics[width=0.6\textwidth]{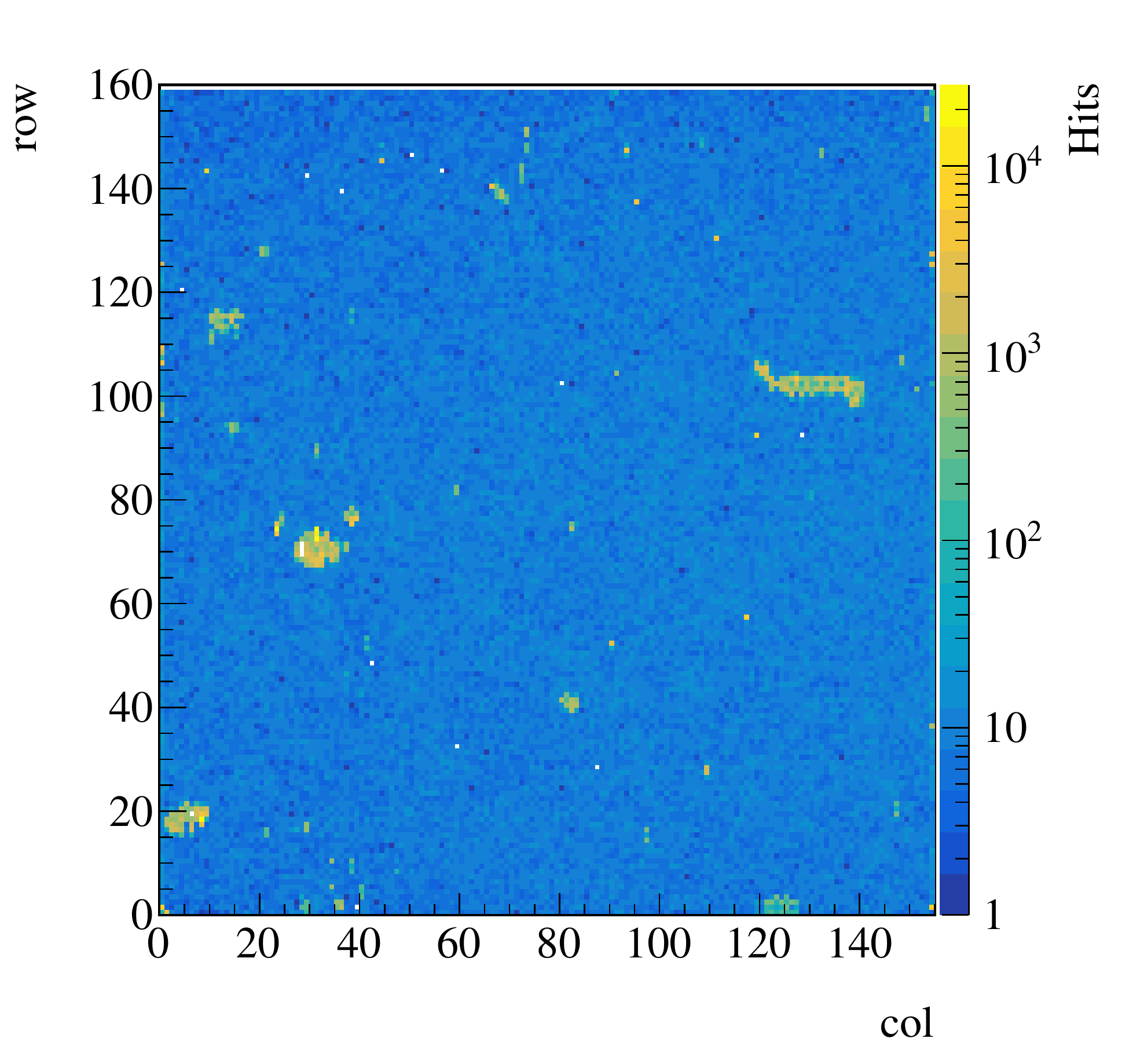}
  \caption{Hit map for sensor A at perpendicular incidence. All 24800 pixels are shown. Noisy regions from bad UBM deposition are masked out from the analysis.}
  \label{fig:hitmap}
\end{figure}

The sensors were operated at room temperature and at a bias voltage of 25V, safely above the full depletion voltage obtained from 
CV measurements (below 10 V, see Figure~\ref{fig:PADIVCV}). Figure~\ref{fig:hitmap} shows the hit map for sensor A at perpendicular incidence. 
All $155 \times 160$ pixels are shown. Apart from the noisy 
regions which are due to a bad UBM deposition, the sensor shows a large area of uniform behavior which is more than acceptable for characterization purposes.

\begin{figure}[!hbt]
  \centering
    \begin{subfigure}{.49\textwidth}
        \includegraphics[height=0.9\textwidth,width=\textwidth]{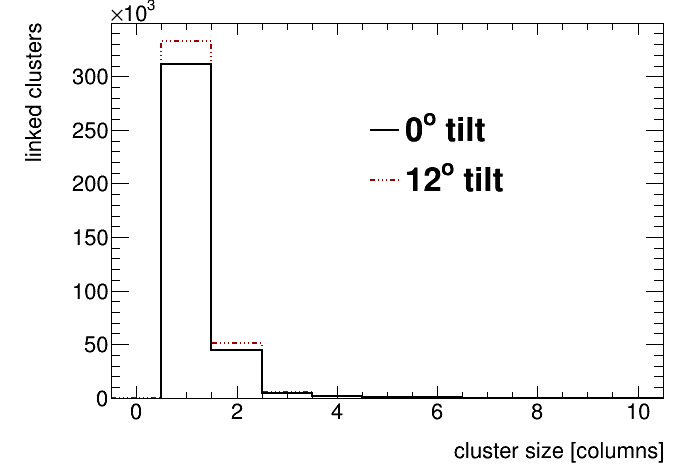}
    \end{subfigure}
    \begin{subfigure}{.49\textwidth}
        \includegraphics[height=0.9\textwidth,width=\textwidth]{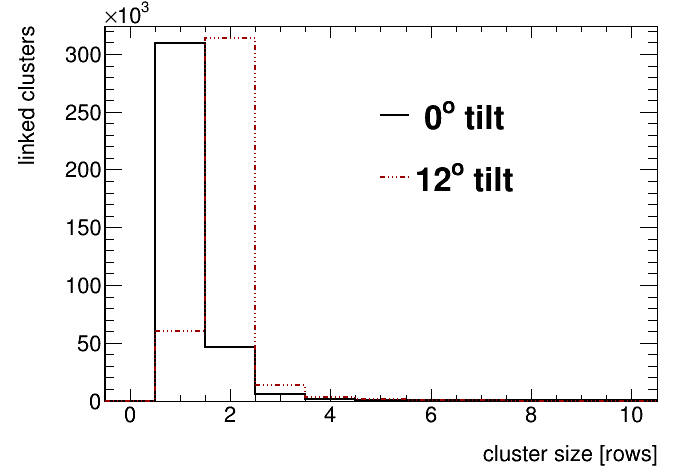}
    \end{subfigure}
    \caption{Distribution of the number of cluster columns (left) and rows (right) for perpendicular and for 12\degr\ incidence.}
    \label{fig:clussize}
\end{figure}

The track incidence angle is measured with respect to the direction perpendicular to the sensor. High statistics samples were recorded for both 
sensors A and B at perpendicular track incidence (0\degr) and at 12\degr, the arc-tangent of the sensor pitch to width ratio. 
Figure~\ref{fig:clussize} shows the cluster
size distribution in number of cluster columns (sensor X direction) and rows (sensor Y direction). The sensor was oriented such that it rotated around its 
X-axis, and therefore the number of cluster columns peaks at 1 regardless of the angle of incidence, while the most likely number of cluster rows clearly increases
with sensor inclination. 


\subsection{Charge}

\begin{figure}
  \centering
    \includegraphics[width=0.6\textwidth]{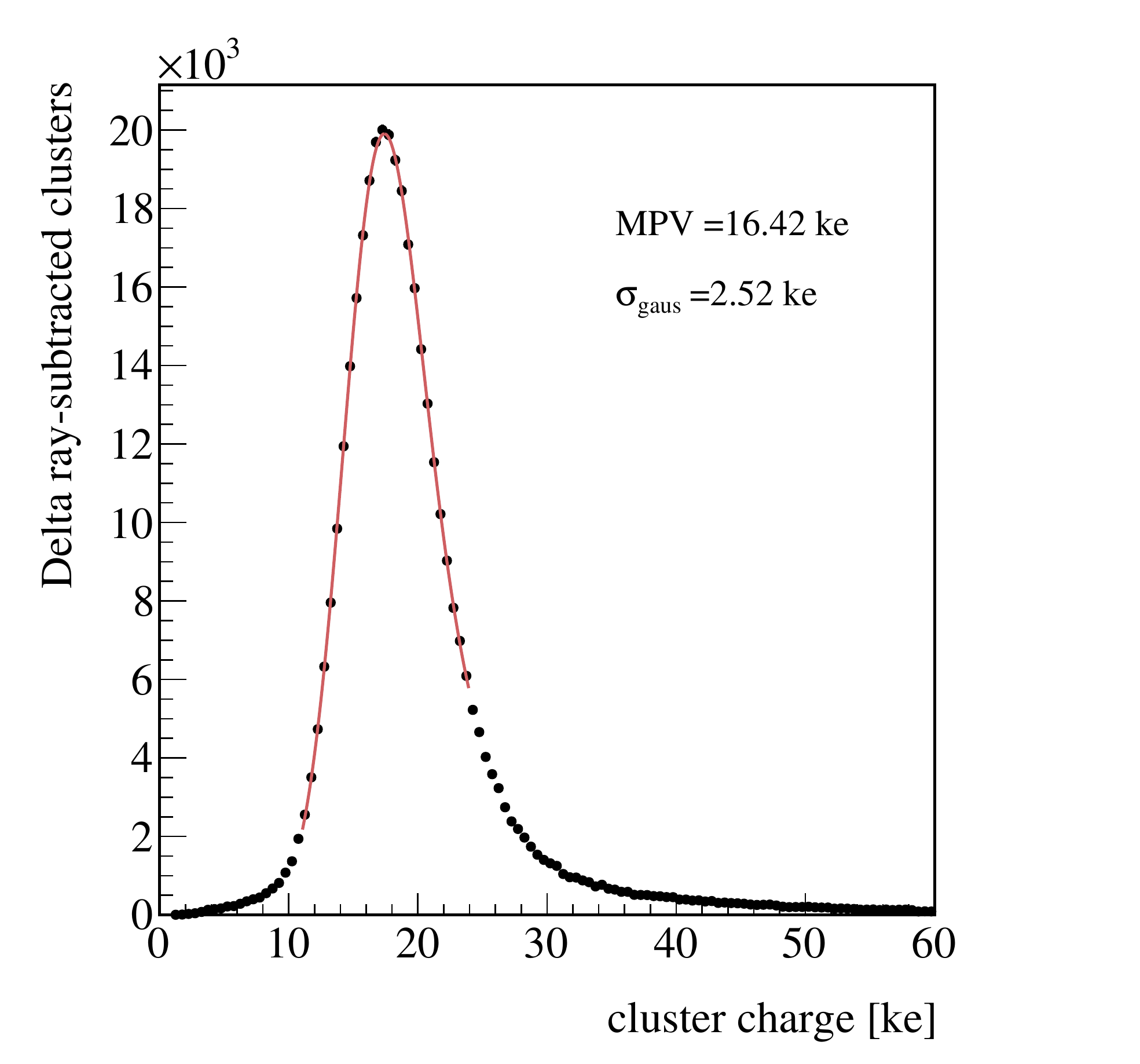}
  \caption{Cluster charge for clusters matched to a track. The charge is normalized to peak at 76 electrons per micron, or 
           17 kilo electrons for a 230 \um\ thick sensor.}
  \label{fig:cluschargelandau}
\end{figure}

\begin{figure}
  \centering
  \begin{subfigure}{.49\textwidth}
      \includegraphics[width=\textwidth]{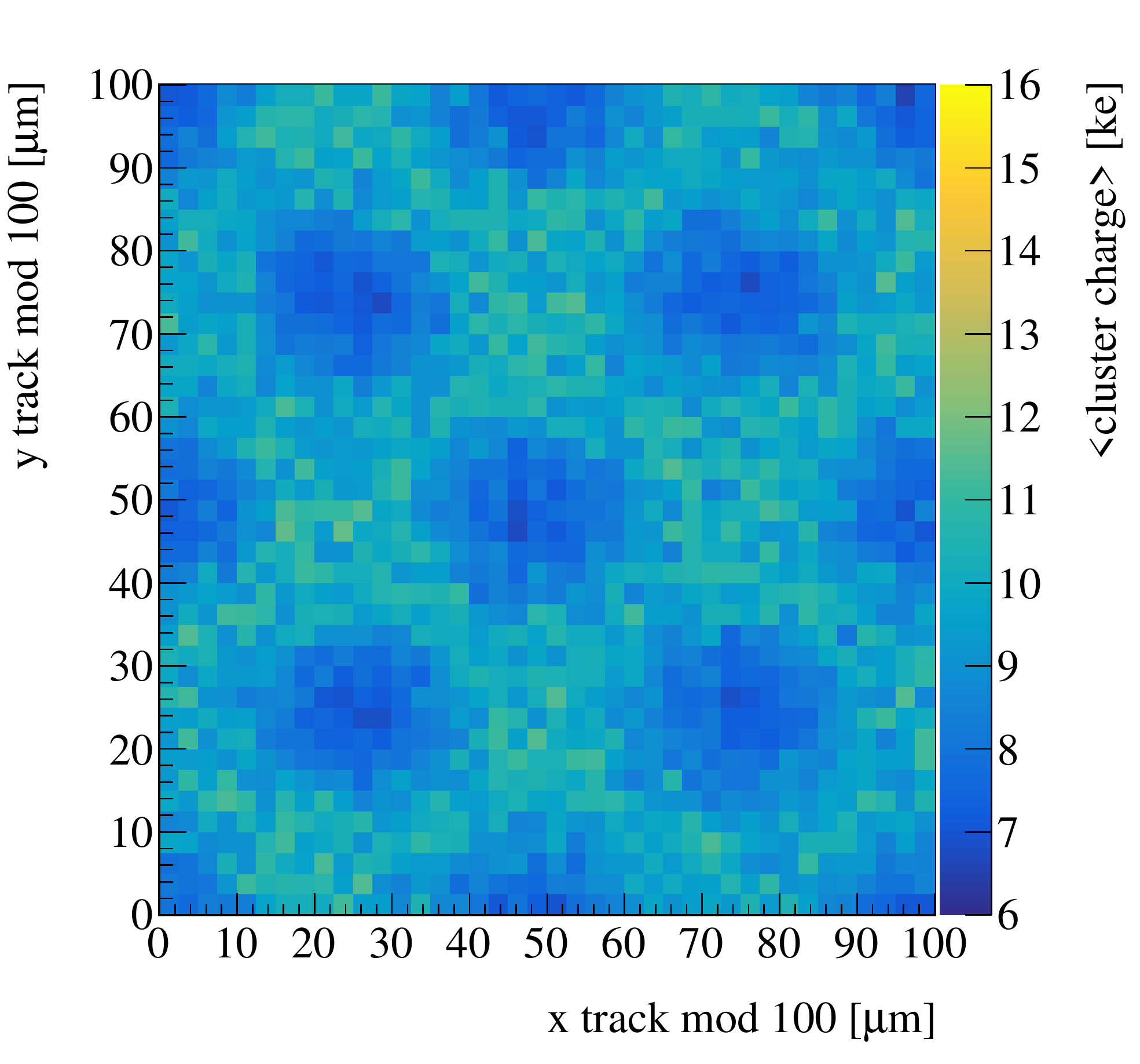}
  \end{subfigure}                                                                                                                             
  \begin{subfigure}{.49\textwidth}
      \includegraphics[width=\linewidth]{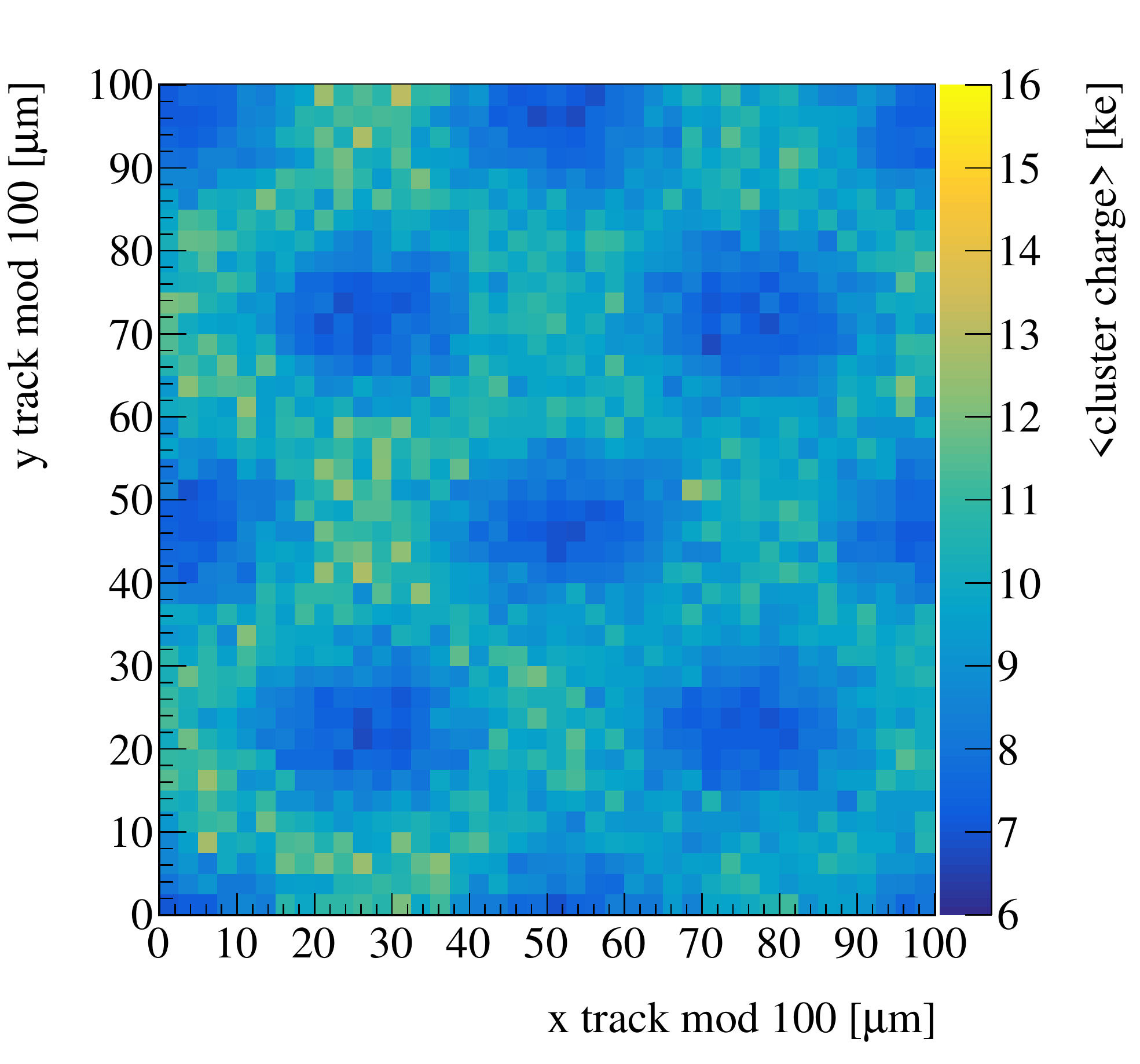}
  \end{subfigure}
  \caption{Cluster charge at perpendicular incidence as a function of track position in an array of $2 \times 2$ pixels for sensor A (left)
           and B (right).}
  \label{fig:clusQmod0}
\end{figure}

\begin{figure}[!htpb]
  \centering
  \begin{subfigure}{.49\textwidth}
      \includegraphics[width=\textwidth]{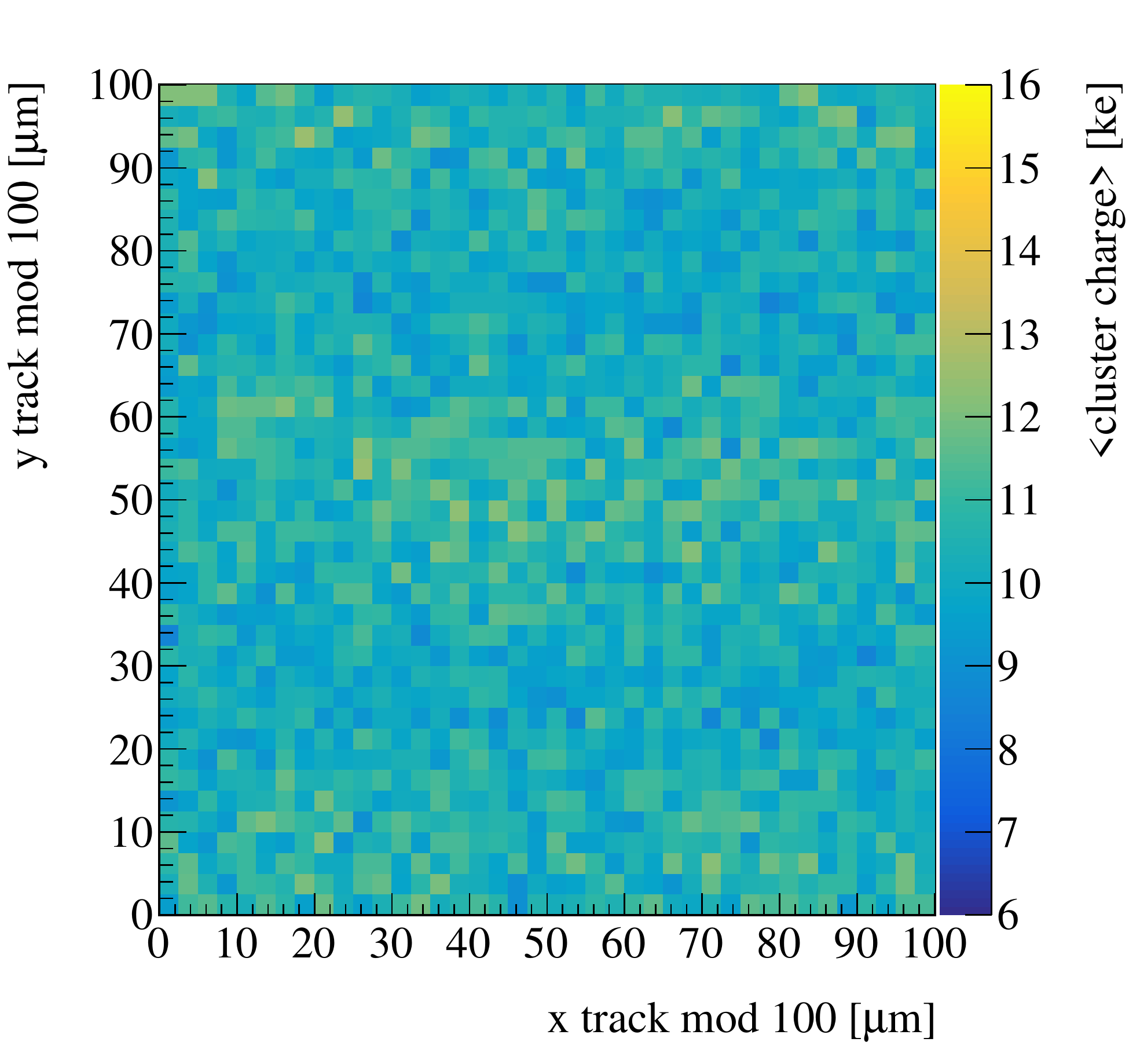}
  \end{subfigure}
    \begin{subfigure}{.5\textwidth}
        \includegraphics[width=\textwidth]{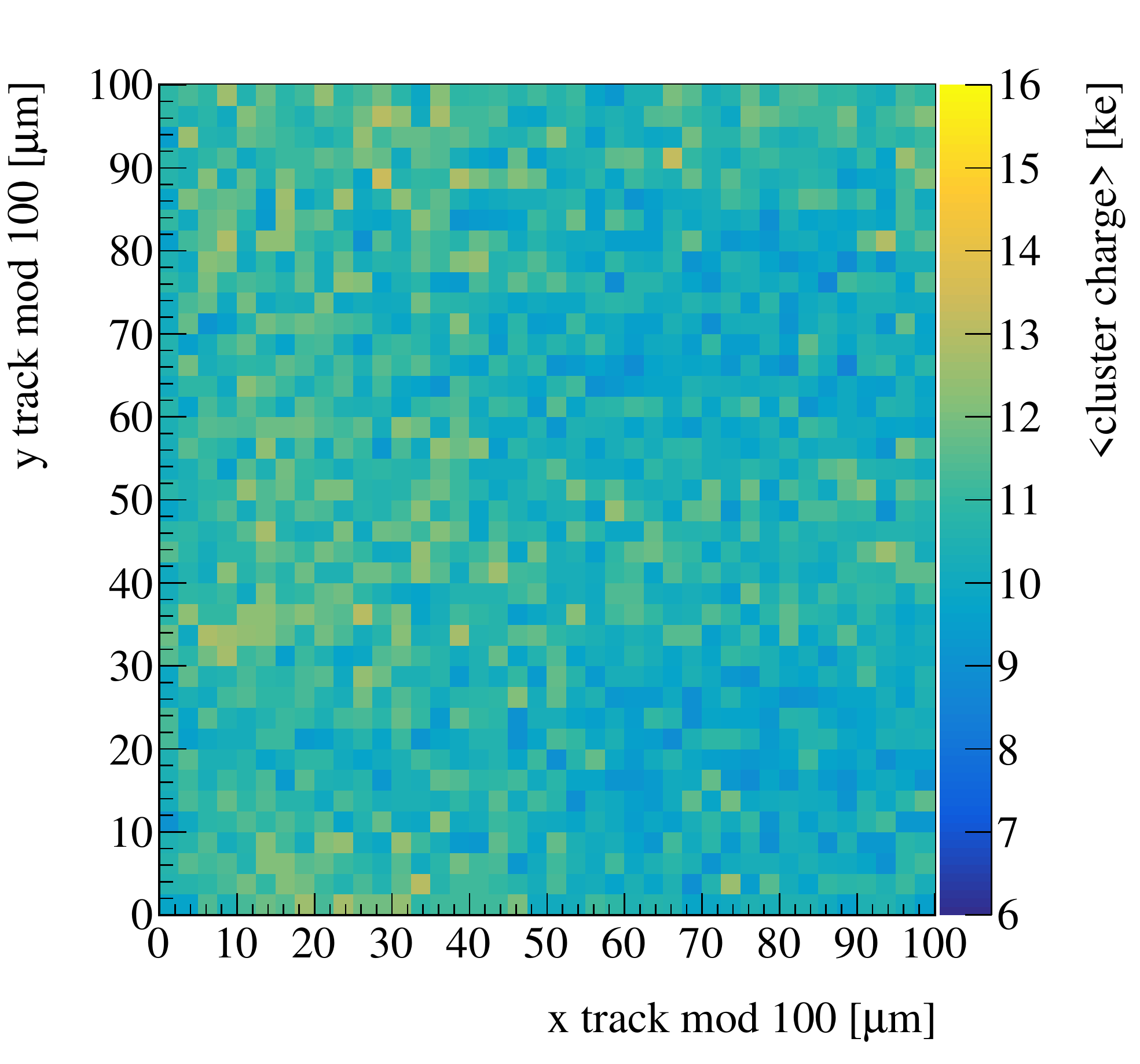}
    \end{subfigure}
    \caption{Cluster charge at 12\degr\ incidence as a function of track position in an array of 
    $2 \times 2$ pixels for sensor A (left) and B (right).}
    \label{fig:clusQmod12}
\end{figure}

Figure~\ref{fig:cluschargelandau} shows the calibrated charge distribution for clusters matched to a track for sensor A at perpendicular incidence. 
Recall that the charge is normalized to the expected MIP charge deposition of 76 electrons per micron, or 17 kilo electrons for a 230 \um\ thick sensor.
The charge collection can be studied as a function of pixel cell position. Figure~\ref{fig:clusQmod0} shows the cluster charge as a 
function of track position for sensors A and B at perpendicular incidence in an array of $2 \times 2$  pixel cells. 
It is important to remark that in this figure, and in all subsequent figures showing a four-pixel array, information from all such (good) 
pixel arrays from the entire sensor is mapped together into a single pixel array in order to obtain reasonably high statistics. 
The $x$ and $y$ axis in these figures are the track coordinates at the plane of the DUT modulo 100 \um. At perpendicular incidence, 
the pixel cell structure can be deduced from the regions of lower charge collection, which correspond to the positions of the union and ohmic columns. 
The structure observed in this figure (and similarly in the efficiency figures below) should be compared to the 
structure of Figure~\ref{fig:3Dsketch} (right) delimited by the nine ohmic columns shown, defining a 2x2 pixel array.
The charge collection becomes uniform along the pixel 
cell for non-perpendicular incidence. Figure~\ref{fig:clusQmod12} shows the corresponding charge distributions for 12\degr\ incidence.

\subsection{Efficiency}
\begin{figure}[!htpb]
  \centering
    \begin{subfigure}{.49\textwidth}
        \includegraphics[width=\textwidth]{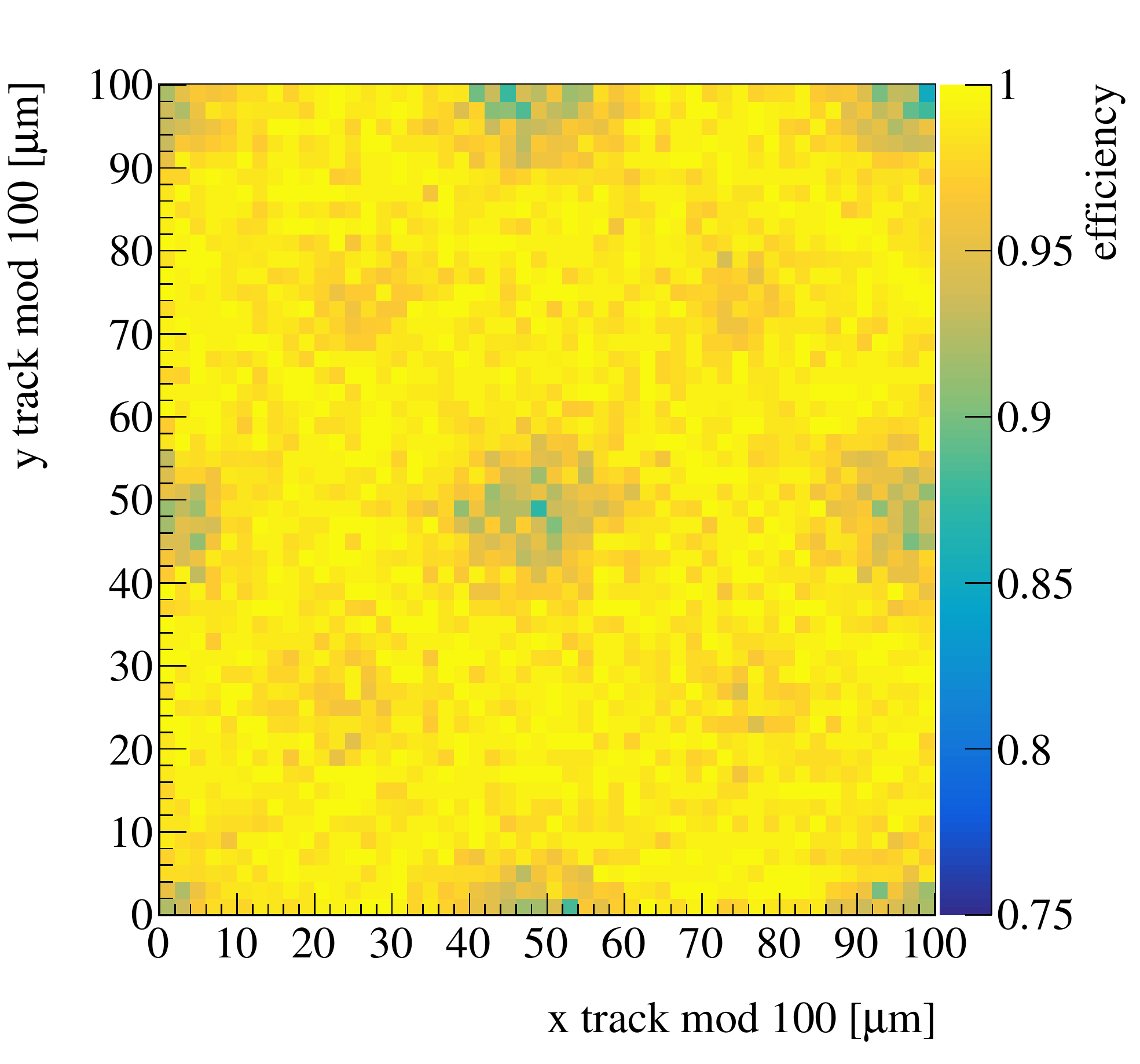}
    \end{subfigure}
    \begin{subfigure}{.49\textwidth}
        \includegraphics[width=\textwidth]{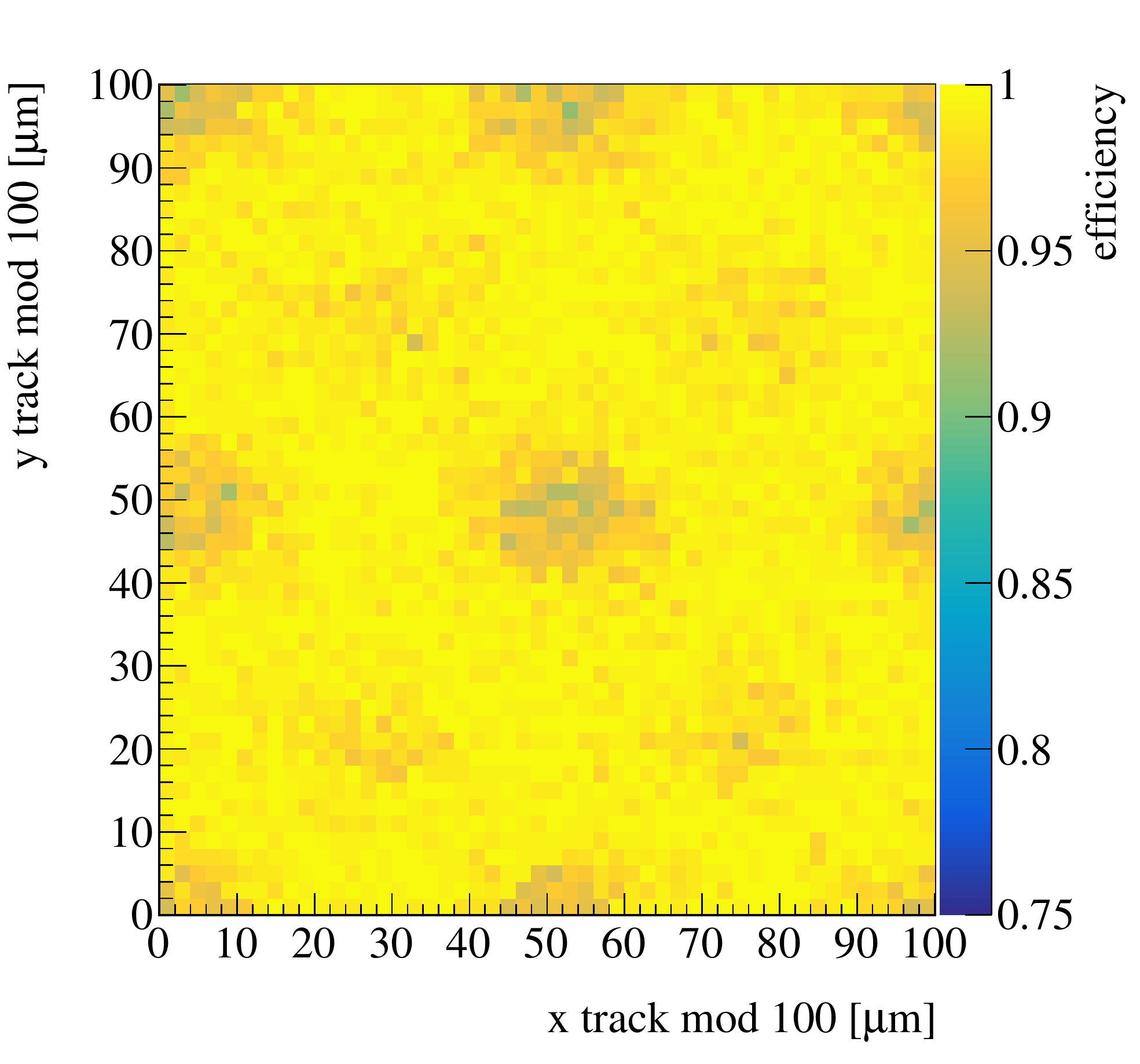}
    \end{subfigure}
    \caption{Sensor tracking efficiency at perpendicular incidence as a function of track position in
    an array of $2 \times 2$ pixels for sensor A (left) and B (right).}
    \label{fig:effmod0deg}
\end{figure}

\begin{figure}[!htpb]
  \centering
    \begin{subfigure}{.49\textwidth}
    \includegraphics[width=\textwidth]{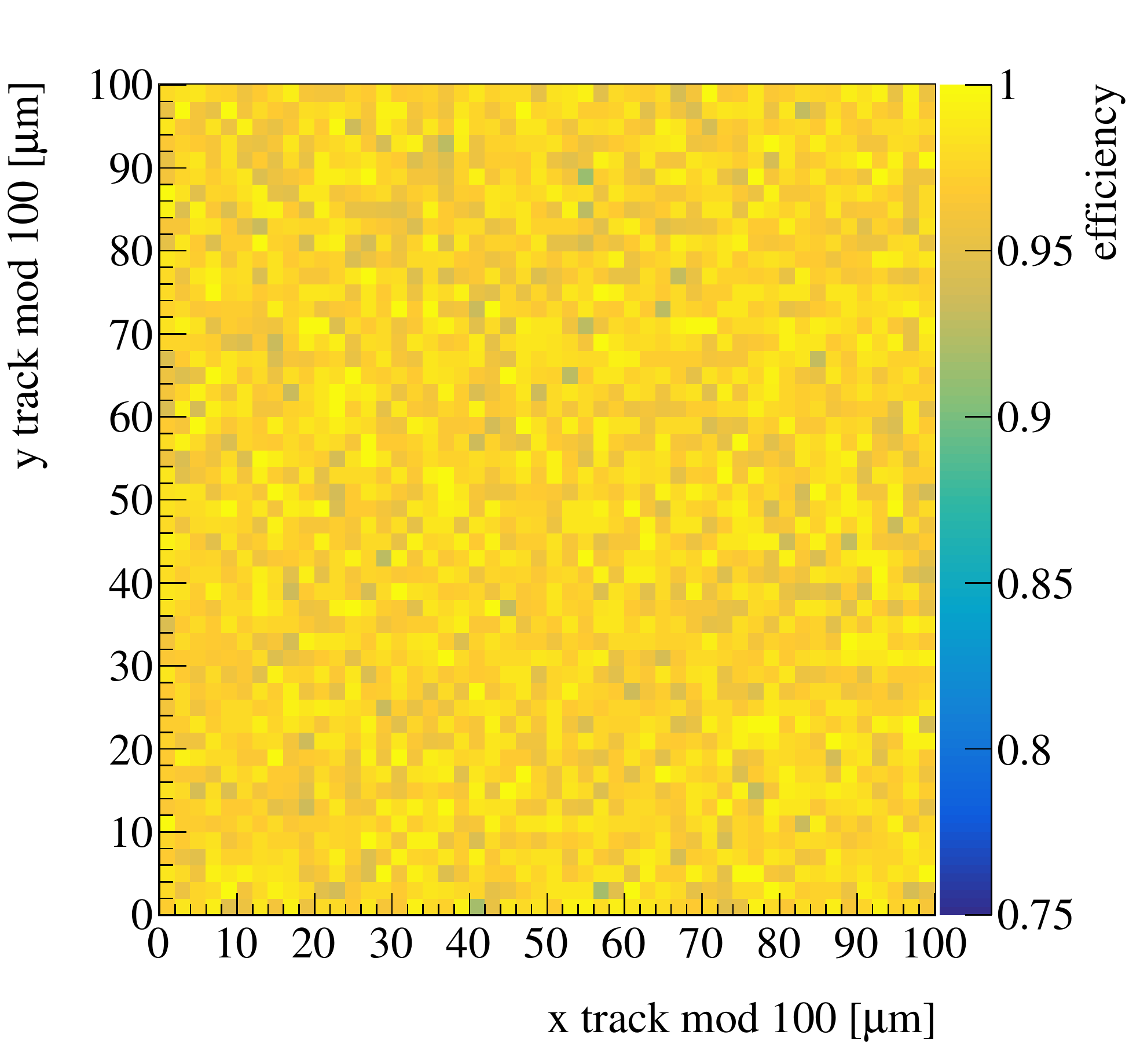}
    \end{subfigure}
    \begin{subfigure}{.49\textwidth}
        \includegraphics[width=\textwidth]{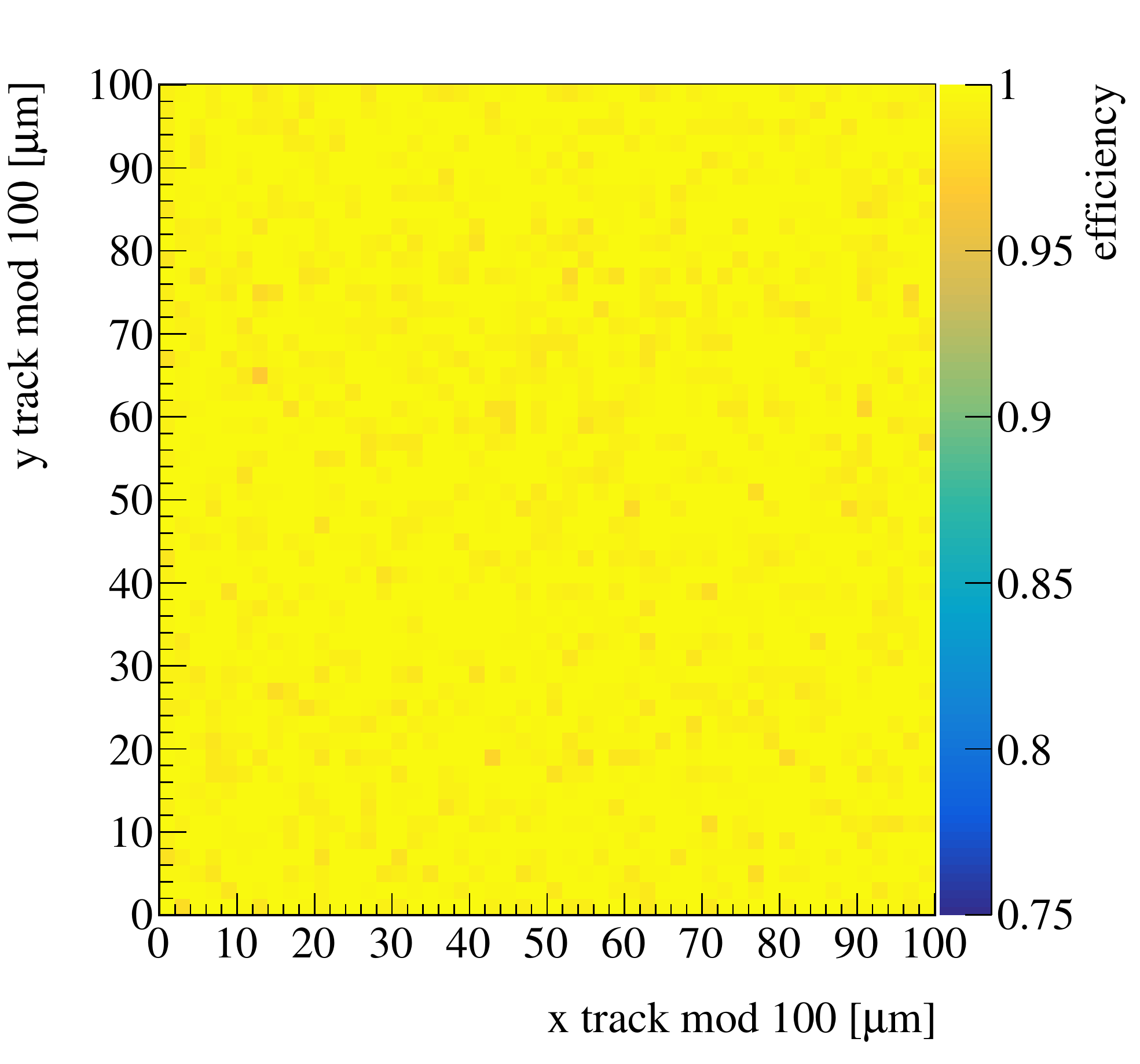}
    \end{subfigure}                                                                                                                             
    \caption{Sensor tracking efficiency at 12\degr\ incidence as a function of track position in an array
    of $2 \times 2$ pixels for sensor A (left) and B (right).}
    \label{fig:effmod12deg}
\end{figure}

The hit detection efficiency is defined as the fraction of telescope tracks matched to the reference sensor which also show a hit matched to the DUT.
Telescope tracks are selected by requiring hits on all six planes of the telescope. In addition, two tracks are formed by using the set of three planes before
and after the DUT. The distance between the position of these two tracks extrapolated to the plane of the DUT is required to be smaller than 100 \um. 
A track is matched to the reference sensor if there is a hit in the sensor at a distance smaller than twice its binary resolution. Matched tracks are also required
to be isolated by imposing the condition that no additional tracks be found within a radius of 300 \um\ from the matching hit. Finally, tracks passing these 
selection criteria are matched to the DUT if a DUT his is found at a distance smaller than 6 times its binary resolution from the track projection at the DUT plane.
The hit definition uses a threshold of $\Delta PH > 24$ ADC counts compared to the neighbouring pixels as discussed in Sec. 3.
Figure~\ref{fig:effmod0deg} shows the sensor efficiency at perpendicular incidence for both sensors as a function of track position in an array of $2 \times 2$ 
pixel cells. At perpendicular incidence, 
the efficiency is very close to 100\% except in the regions of the union (n-type) columns, where it drops to approximately 94\%, and the ohmic columns (p-type) 
where it is approximately 97\%. This slight loss of efficiency is recovered when tilting the sensor. Figure~\ref{fig:effmod12deg} shows the corresponding
efficiency for an incidence angle of 12\degr, where it can be seen that the efficiency is essentially constant and close to 100\% throughout the sensor.

\begin{figure}
  \centering
    \includegraphics[width=0.6\textwidth]{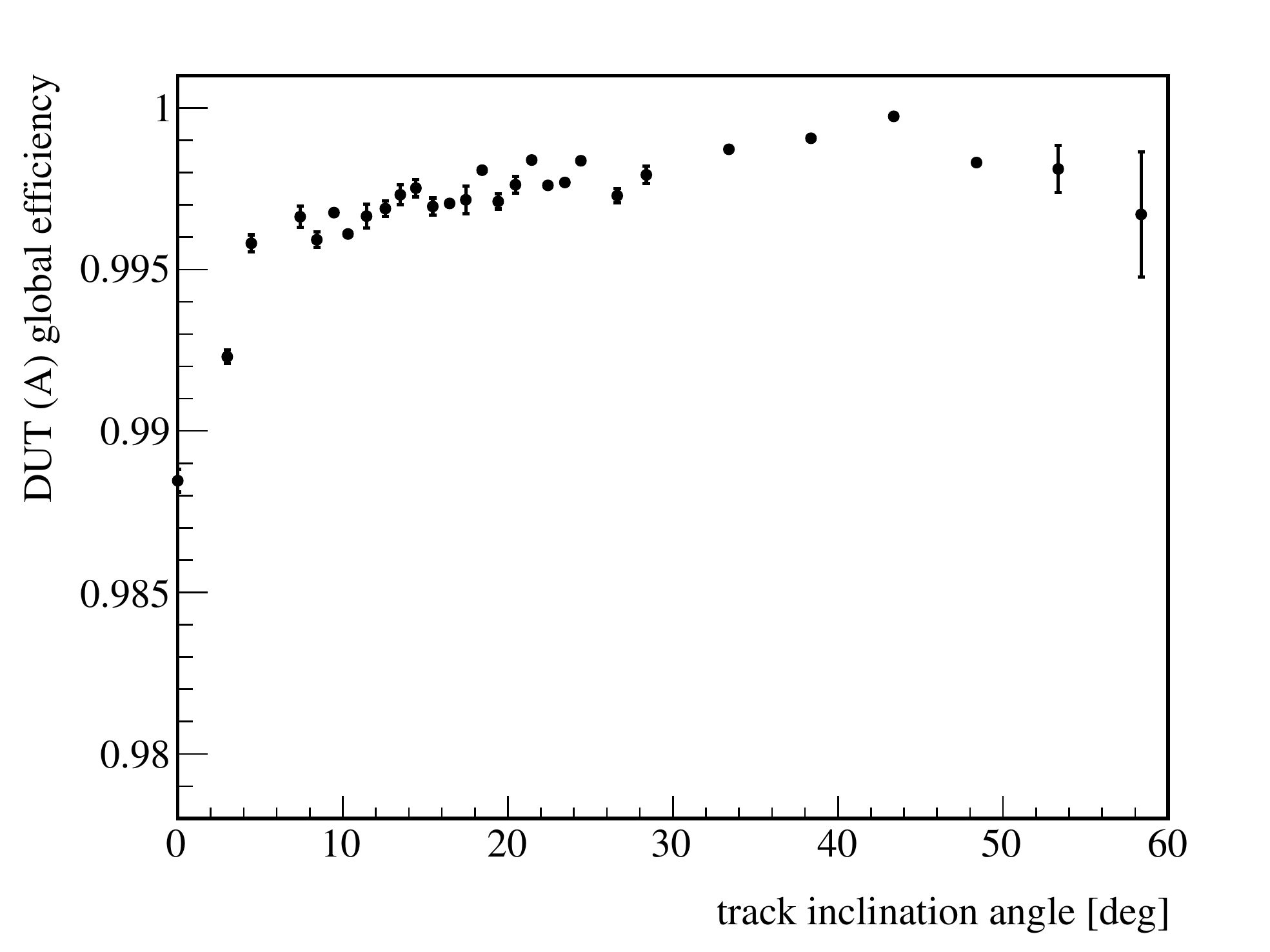}
    \caption{Overall sensor efficiency as a function of track incidence angle}
  \label{fig:effvsangle}
\end{figure}

The overall efficiency, integrated over the entire sensor, was studied as a function of incidence angle for sensor A. The results are summarized in 
Figure~\ref{fig:effvsangle}. For incidence angles above 5\degr, the overall efficiency remains above 99.5\%. 

\subsection{Charge sharing}
\begin{figure}[!htpb]
    \centering
    \begin{subfigure}{.49\textwidth}
        \includegraphics[width=\textwidth]{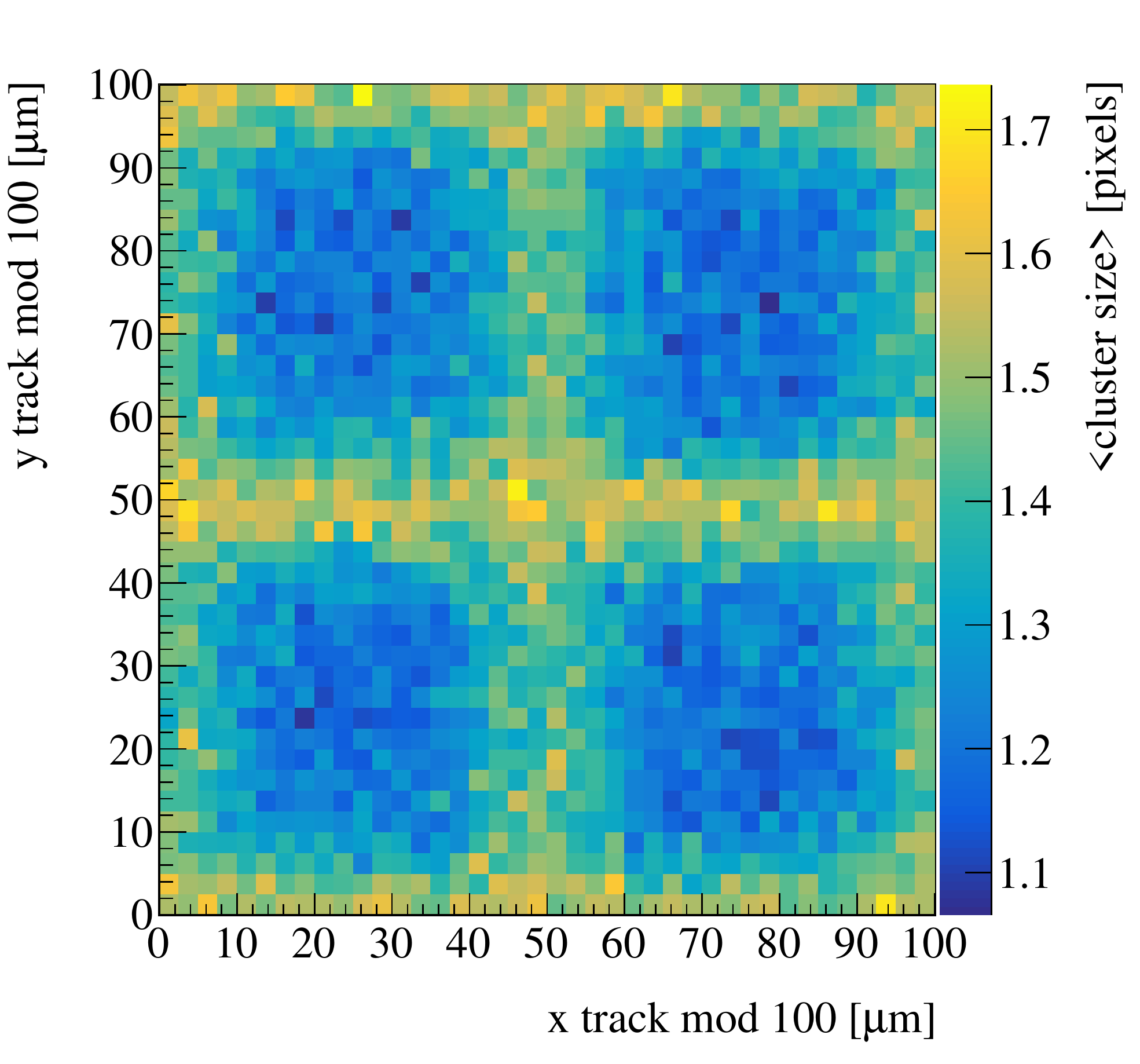}
    \end{subfigure}
    \begin{subfigure}{.49\textwidth}                                                                                                             
        \includegraphics[width=\textwidth]{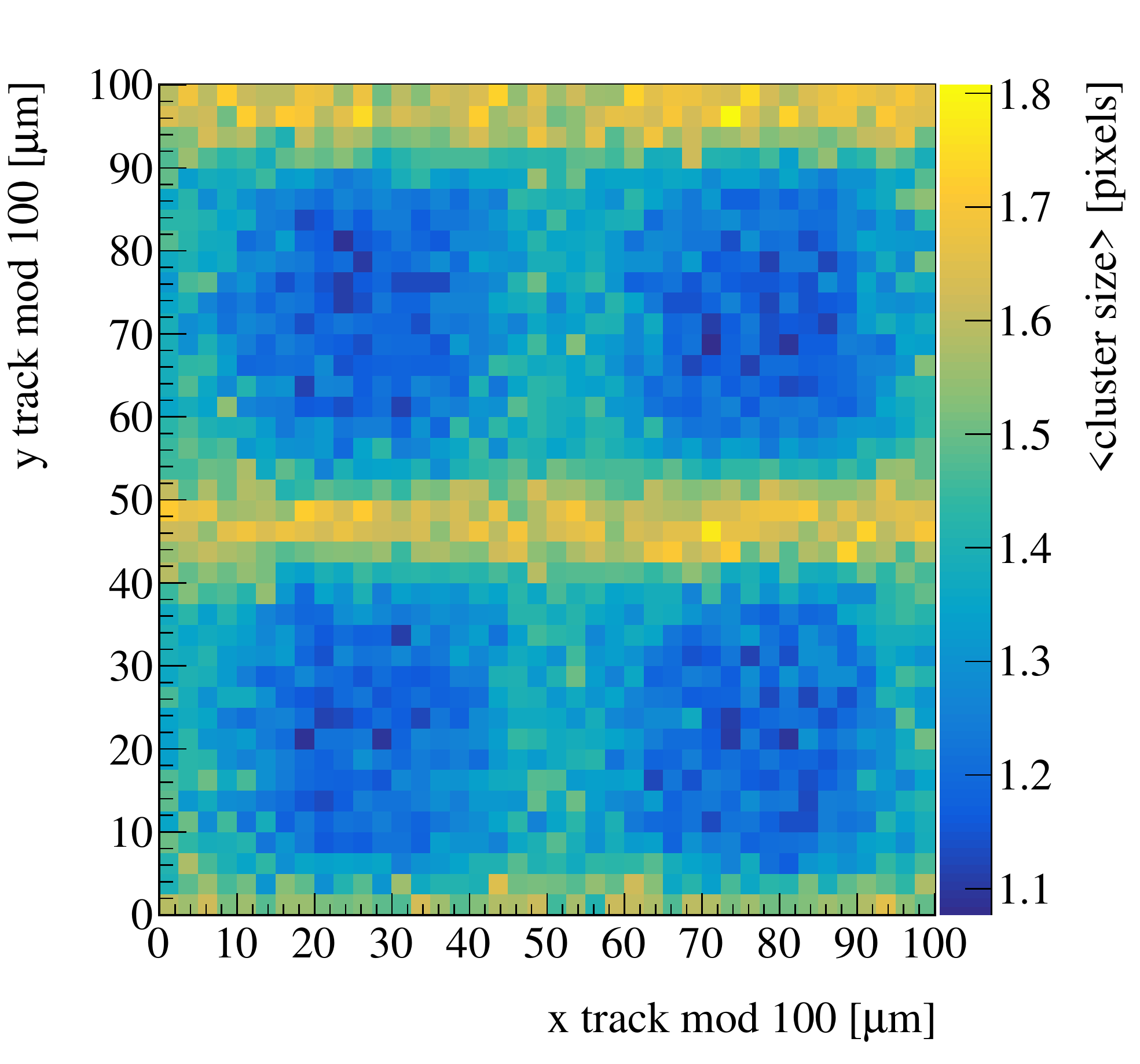}
    \end{subfigure}
    \caption{Mean cluster size at perpendicular incidence as a function of track position in 
    an array of $2 \times 2$ pixels for sensor A (left) and B (right).}
    \label{fig:clussizemod0deg}
\end{figure}

\begin{figure}[!htpb]
    \centering
    \begin{subfigure}{.49\textwidth}
        \includegraphics[width=\textwidth]{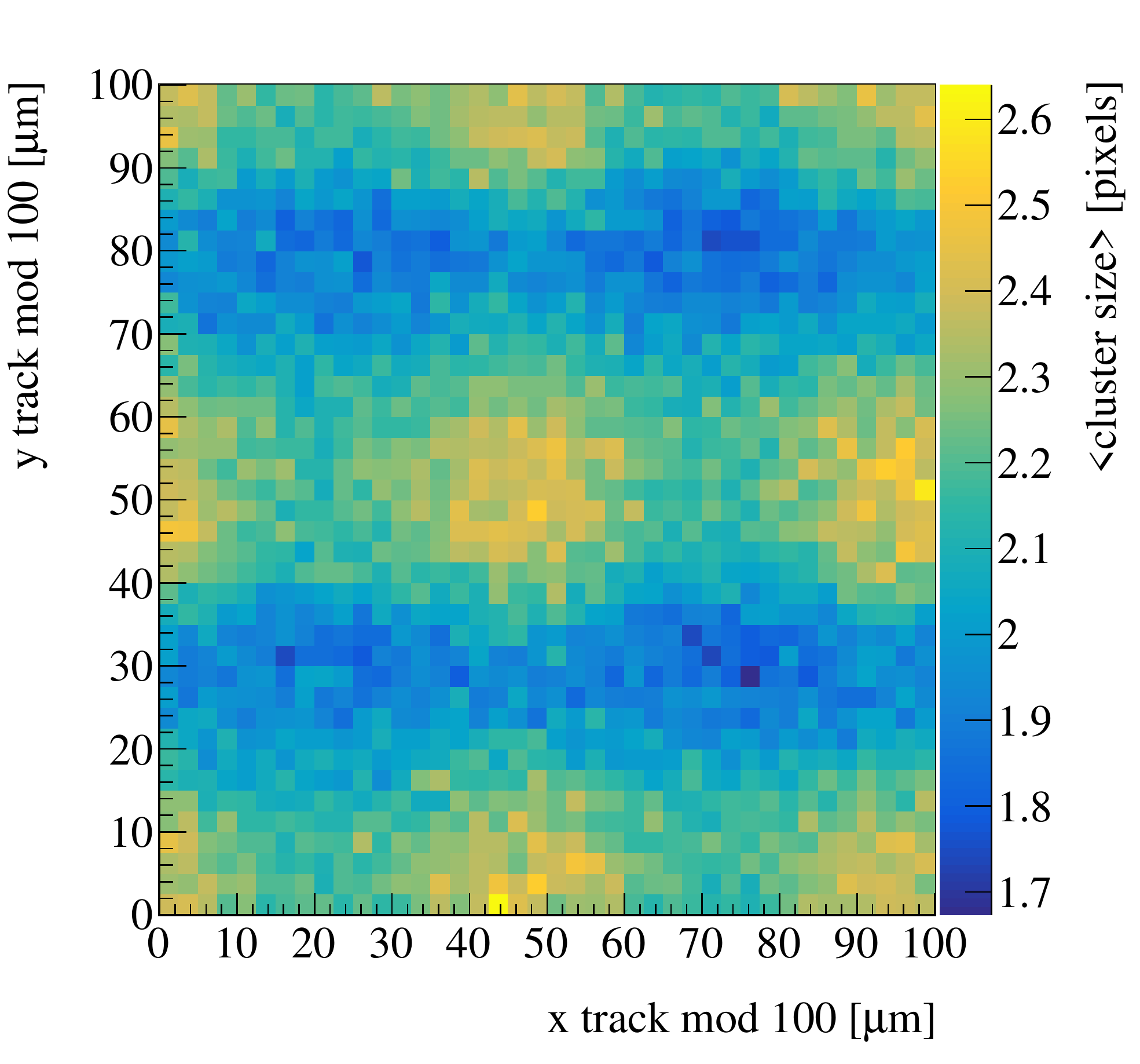}
    \end{subfigure}                      
    \begin{subfigure}{.49\textwidth} 
        \includegraphics[width=\textwidth]{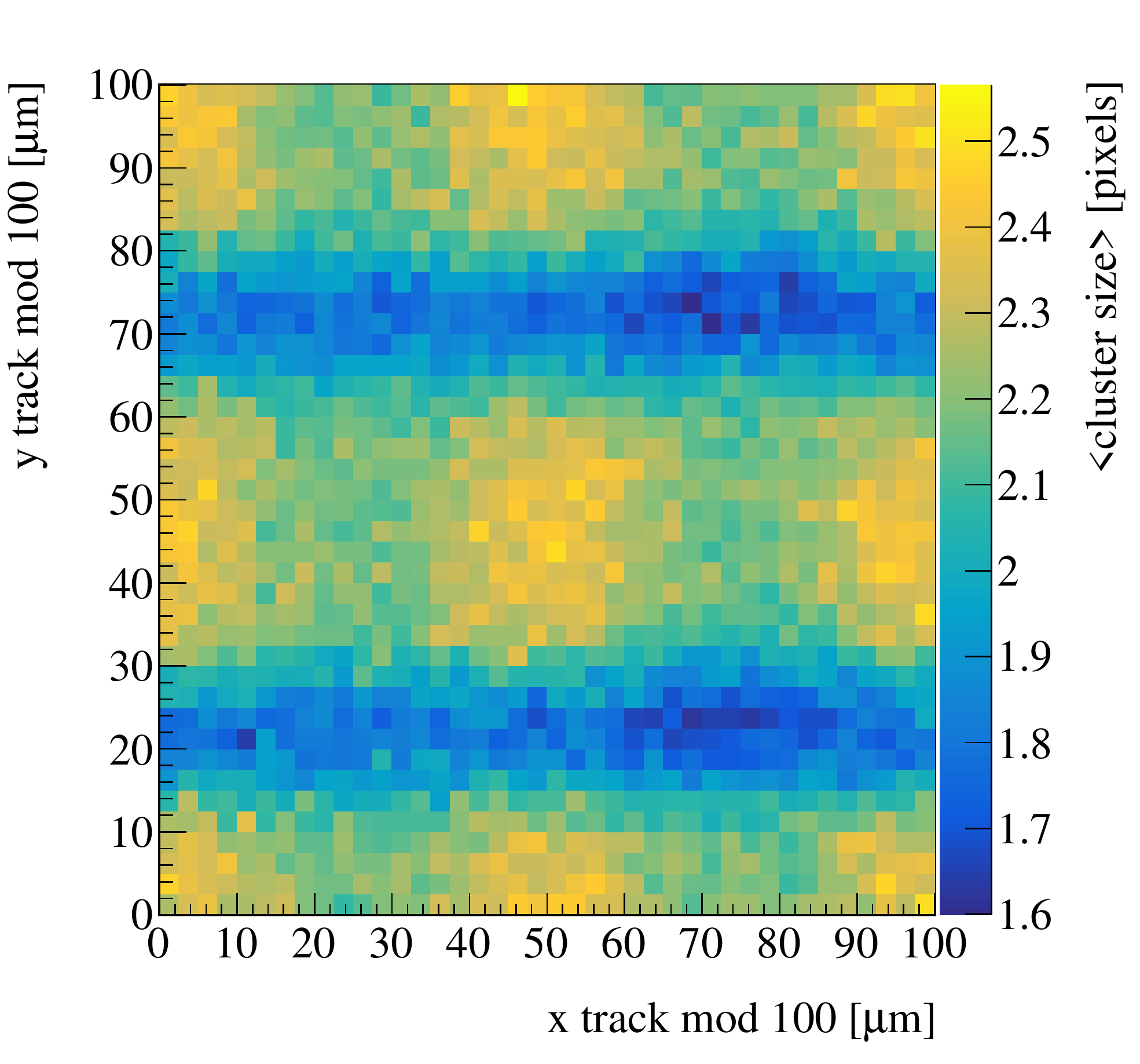}
    \end{subfigure}
    \caption{Mean cluster size at 12\degr\ incidence as a function of track position in 
    an array of $2 \times 2$ pixels for sensor A (left) and B (right).}
    \label{fig:clussizemod12deg}
\end{figure}

The sharing of pixel cluster charge between adjacent pixels helps to improve the spatial resolution with respect to the binary resolution. Charge sharing is 
expected to be largest near pixel boundaries. Figure~\ref{fig:clussizemod0deg} shows the mean cluster size at perpendicular incidence. As expected, the 
mean cluster size is close to unity near the pixel cell centers, which are dominated by single-pixel clusters. The mean cluster size rises significantly above 1 
near pixel boundaries where clusters of 2 and 3 pixels make a significant contribution. Figure~\ref{fig:clussizemod12deg} shows the corresponding mean cluster
size for 12\degr\ incidence angle. As expected, the mean cluster size increases significantly over the entire sensor (notice the change of scale), and the 
region of lower charge sharing shrinks closer to the pixel cell center along the tilted direction (Y).  

\subsection{Spatial resolution}
\begin{figure}
  \centering
    \includegraphics[width=0.6\textwidth]{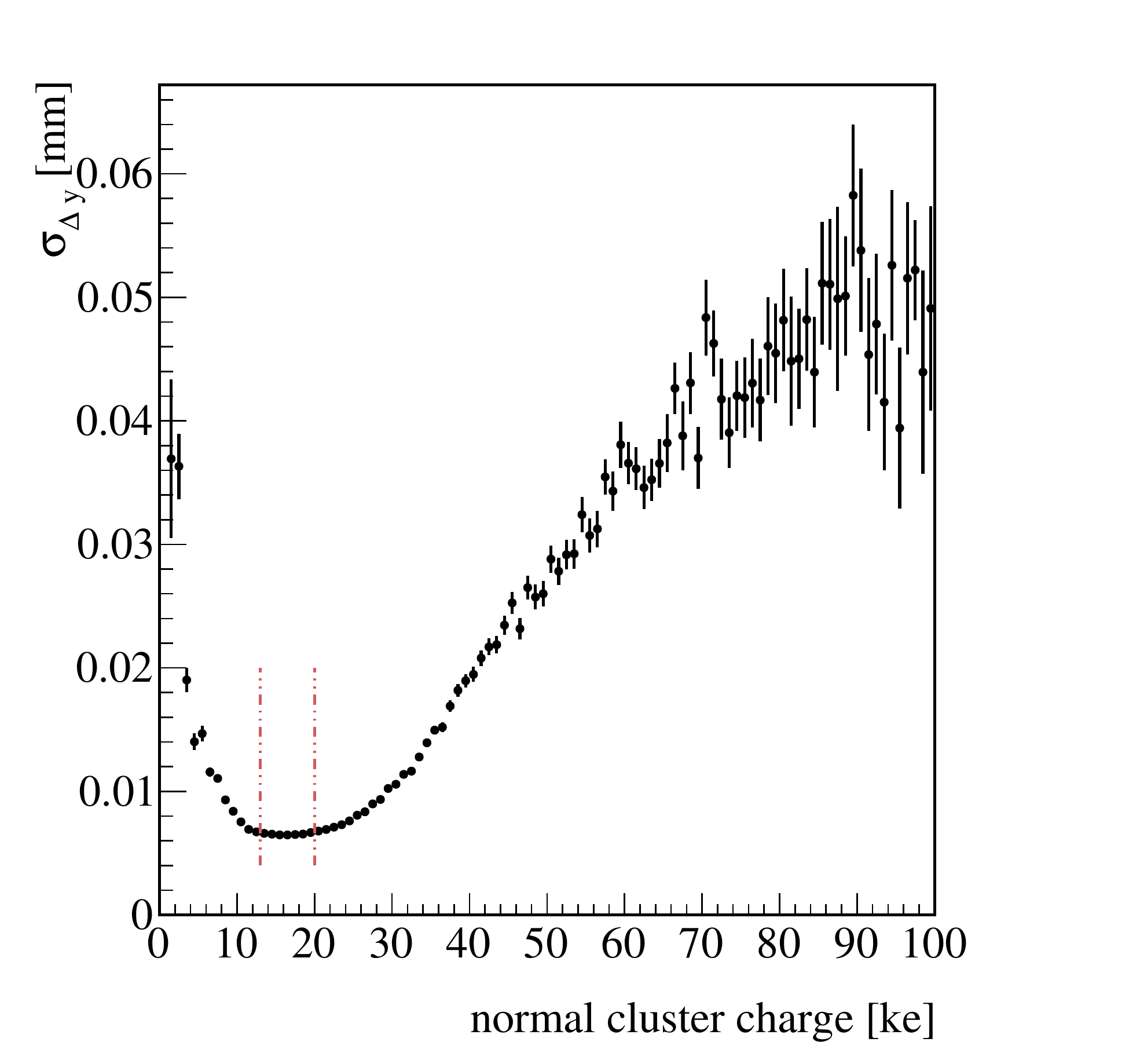}
  \caption{Residual width as a function of cluster charge. Only clusters with charge in the region inside the dashed lines are used for the DUT resolution
           measurement. The resolution deterioration at large charge is due to the influence of delta rays.}
  \label{fig:resvsQ}
\end{figure}

In order to determine the sensor resolution, the DUT is placed as close as physically possible to the upstream telescope triplet. The triplet
track is extrapolated to the plane of the DUT taking into account multiple scattering by using the General Broken Lines (GBL) formalism~\cite{ref:GBL} 
and the distance to the DUT hit position is computed. The width of this residual distribution has a contribution from the telescope 
resolution, which is known, and from the intrinsic DUT resolution, which is therefore calculated as: 
\begin{equation}
\label{eq:dut_resolution}
\sigma_{DUT} = \sqrt{(\sigma_{\Delta y})^2 - (\sigma_{telescope})^2}
\end{equation}
Only DUT clusters matched to a telescope track and to the reference sensor are used. 
In addition, only clusters which have a charge close to the Landau most probable value are used. This cut in charge rejects delta rays 
produced in the sensor which would severely affect the position resolution, since delta electrons produced at the point of incidence typically travel 
a long distance inside the sensor, producing very large clusters of large charge and with a centroid far away from the track incidence point. 
Figure~\ref{fig:resvsQ} shows the residual width as a function of cluster charge, where the deterioration at large charge due to delta rays can be seen. 

\begin{figure}[!htpb]
  \centering
    \includegraphics[width=0.6\textwidth]{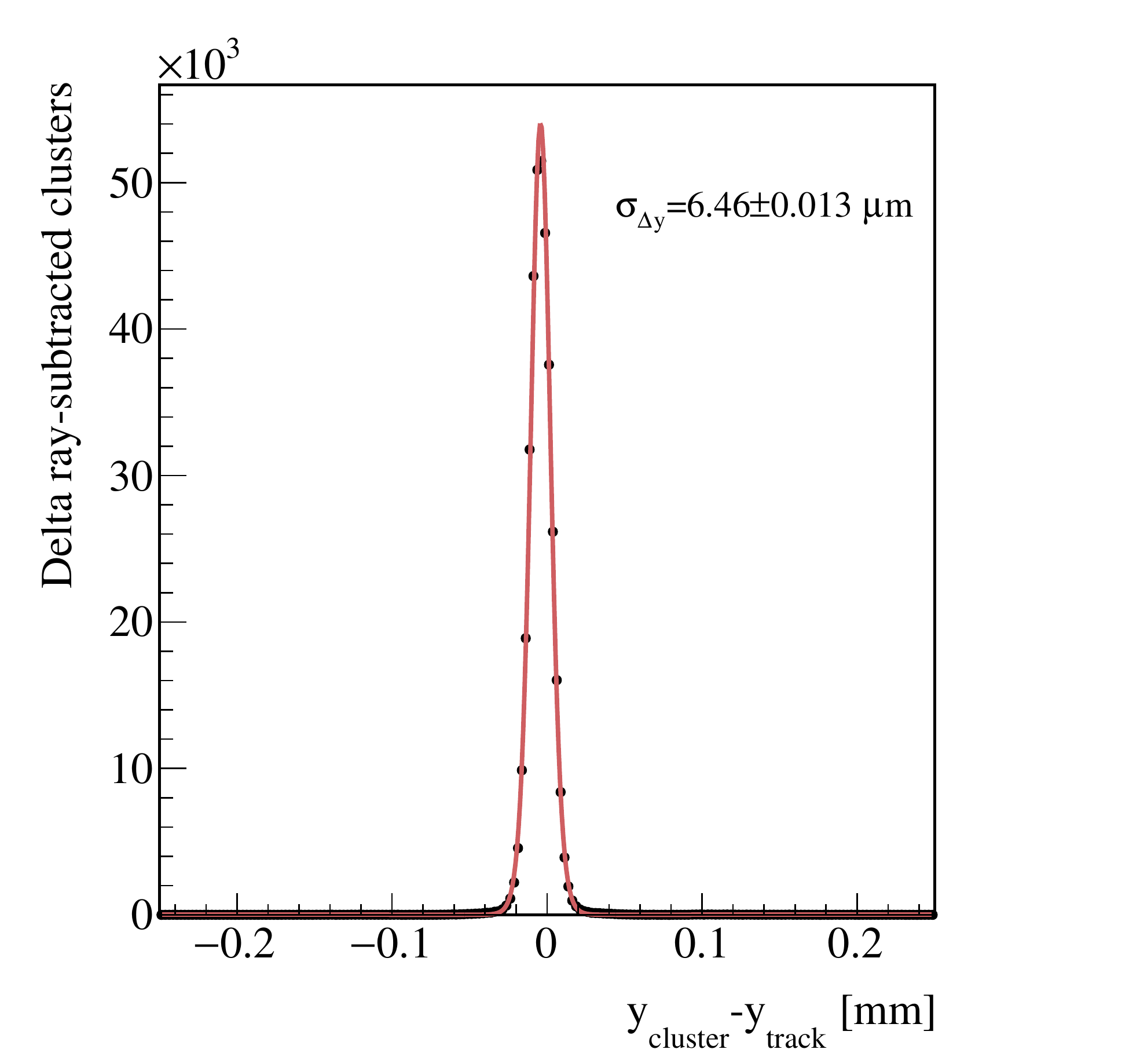}
  \caption{Residual distribution for 12\degr\  incidence for clusters matched to a telescope track after delta ray rejection. The width has a contribution
           from the telescope resolution and from the DUT resolution.}
  \label{fig:dutres}
\end{figure}

The optimal DUT resolution is calculated using clusters from 12\degr\ incidence tracks after rejecting delta rays. The residual distribution
between the telescope track projection and the cluster center has a width of 6.46 \um, as shown in Figure~\ref{fig:dutres}. After subtracting in quadrature the 
telescope resolution evaluated with the GBL formalism, the DUT resolution at 12\degr\ incidence is found to be 3 \um.

\begin{figure}
  \centering
    \includegraphics[width=0.6\textwidth]{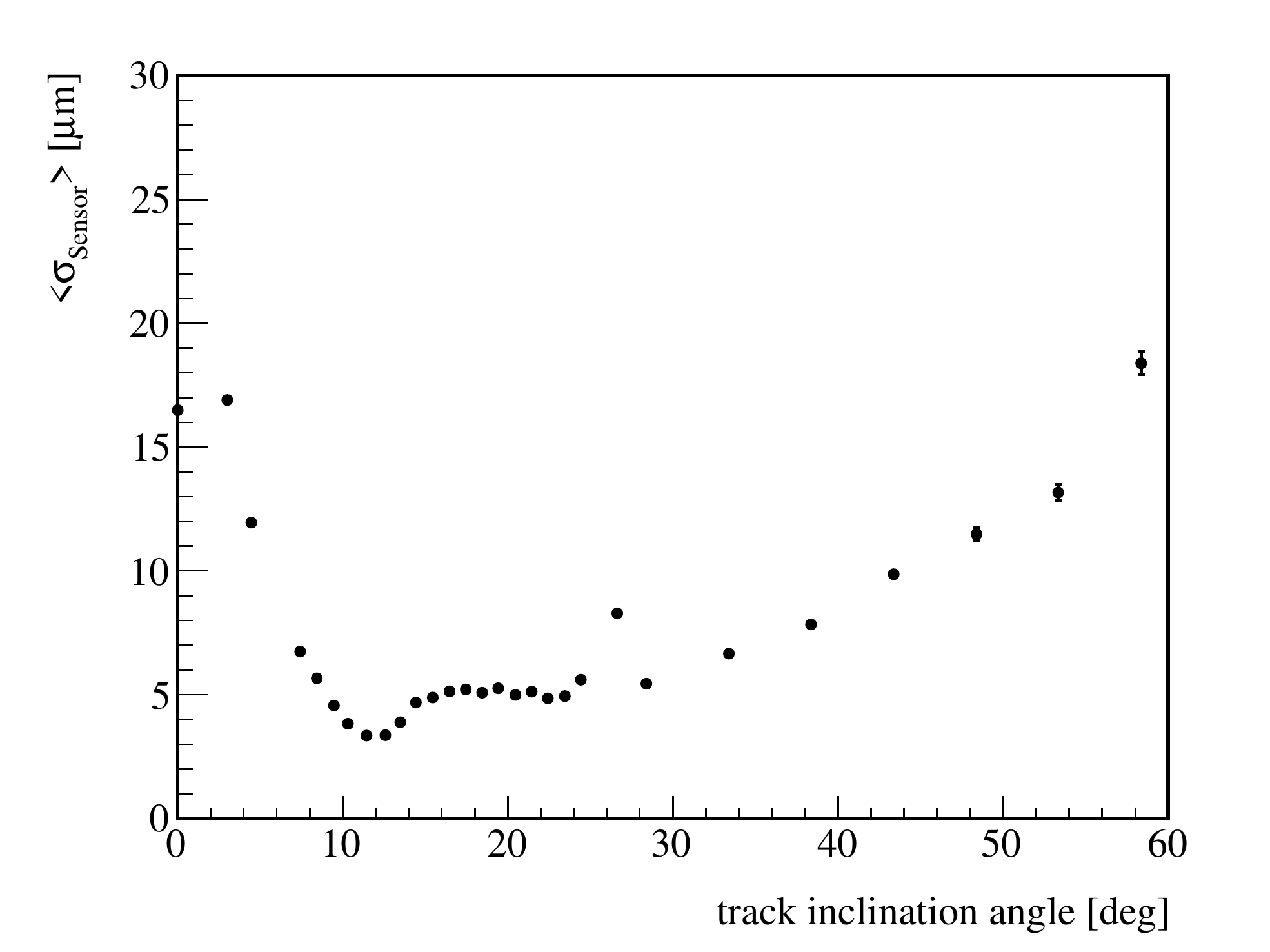}
  \caption{DUT resolution as a function of track incidence angle.}
  \label{fig:resvsangle}
\end{figure}

The resolution has been studied as a function of incidence angle. The resolution is expected to be close to the binary resolution at perpendicular incidence 
(14.4 \um\ for a pitch of 50 \um), and improve with charge sharing as the cluster size increases, reaching an optimal value at the arc-tangent of the pitch to 
thickness ratio (12\degr\ for these sensors). Figure~\ref{fig:resvsangle} shows the DUT resolution as a function of incidence angle. 

\section{Conclusions}
For the first time, small-pitch 3D pixel sensors of cell size \pitch{50}{50} have been characterized using a small-pitch readout chip, the ROC4SENS.
The sensor response to minimum ionizing radiation was studied at a test beam of 5.6 GeV electrons at DESY. The sensors show good performance in terms of
efficiency, charge sharing, and hit position resolution. For incidence angles larger than 5 degrees (with respect to perpendicular incidence), 
the single hit efficiency is found to be larger than 99.5\%, and a hit position resolution between 3 and 4 \um\ at optimal incidence angle is achieved.
The unirradiated sensors were operated at 25 V, safely above the full depletion voltage which is below 10 V, yet considerably lower than typical planar 
sensor operating voltages. In a future characterization, we plan to irradiate these sensors with protons to assess their robustness to irradiation and
therefore their adequacy for the high luminosity upgrades of the LHC vertex detectors. A recent study~\cite{ref:3Dirrad} of small pitch 3D pixels 
irradiated up to a fluence of $3 \times 10^{16}$ \Neq\ shows that the 3D technology is suitable for the inner layers of HL-LHC vertex detectors.  

\section{Acknowledgments}

This project has received funding from the European Union’s Horizon 2020 research and innovation programme under grant agreement No 654168, 
and from the Spanish MINECO ministry, under grants FPA2014-55295-C3-1-R and FPA2015-71292-C2-1-P.
We thank DESY for their test-beam infrastructure and support, and the PSI team involved in the ROC4SENS development. 

\bibliography{R4S3D-arXiv}   

\end{document}